\documentclass[twoside]{article}
\usepackage{epsbox}

\begin{document}

\centerline{\Large \bf 
Detailed Analysis of the Three Quark Potential in SU(3) Lattice QCD 
}

\vspace{0.75cm}

\centerline{
T.T. Takahashi$^1$, H. Suganuma$^2$, Y. Nemoto$^3$ and H. Matsufuru$^4$}

\vspace{0.5cm}

\centerline{\it $^1$ RCNP, Osaka University, Mihogaoka 10-1, 
Ibaraki, Osaka 567-0047, Japan}
\centerline{\it $^2$ Faculty of Science, Tokyo Institute of Technology, 
Ohokayama 2-12-1, Meguro, Tokyo 152-8551, Japan}
\centerline{\it $^3$ Brookhaven National Laboratory, RBRC, Physics 
Dept. 510A, Upton, New York 11973-5000, USA}
\centerline{\it $^4$ YITP, Kyoto University, Kitashirakawa, Sakyo, 
Kyoto 606-8502, Japan}

\vspace{0.3cm}
      
\begin{abstract}
The static three-quark (3Q) potential is studied in detail using SU(3) lattice QCD
with $12^3 \times 24$ at $\beta=5.7$ and $16^3 \times 32$ at $\beta=5.8$, $6.0$
at the quenched level. For more than 300 different patterns of the 3Q systems,
we perform the accurate measurement of the 3Q Wilson loop with the smearing method,
which reduces excited-state contaminations, and present the lattice QCD data of
the 3Q ground-state potential $V_{\rm 3Q}$.
We perform the detailed fit analysis on $V_{\rm 3Q}$ in terms of the Y-ansatz both with the
continuum Coulomb potential and with the lattice Coulomb potential, and find that
the lattice QCD data of the 3Q potential $V_{\rm 3Q}$ are well reproduced within a few \%
deviation by the sum of a constant, the two-body Coulomb term and the three-body linear
confinement term $\sigma_{\rm 3Q} L_{\rm min}$, with $L_{\rm min}$ the minimal
value of the total length of color flux tubes linking the three quarks.
 From the comparison with the Q-$\bar {\rm Q}$ potential, we find a universality of the
string tension as $\sigma_{\rm 3Q} \simeq \sigma_{\rm Q \bar Q}$ and the one-gluon-exchange
result for the Coulomb coefficients as $A_{\rm 3Q} \simeq \frac12 A_{\rm Q \bar Q}$.
We investigate also the several fit analyses with the various ans\"atze:
the Y-ansatz with the Yukawa potential, the $\Delta$-ansatz and
a more general ansatz including the Y and the $\Delta$ ans\"atze in some limits.
All these fit analyses support the Y-ansatz on the confinement part in the
3Q potential $V_{\rm 3Q}$, although $V_{\rm 3Q}$ seems to be approximated by the
$\Delta$-ansatz with $\sigma_\Delta \simeq 0.53 \sigma$.

\end{abstract}

\vspace{0.3cm}

\section{Introduction}
The strong interaction in hadrons or nuclei is fundamentally ruled by quantum chromodynamics
(QCD). In a spirit of the elementary particle physics, it would be desirable to understand
hadron physics and nuclear physics at the quark-gluon level based on QCD.
However, it still remains as a difficult problem to derive even the inter-quark potential
from QCD in the analytic manner, because of the strong-coupling nature of QCD in the
infrared region.

In this decade, the lattice QCD calculation has been adopted as a useful and reliable method
for the nonperturbative analysis of QCD~\cite{C83,R92}. 
In particular, the quark-antiquark
(Q-$\bar{\rm Q}$) potential, which is responsible for the meson properties, has been well
studied using lattice QCD. The Q-$\bar{\rm Q}$ potential
in lattice QCD is well reproduced by a sum of the Coulomb term
due to the perturbative one-gluon-exchange (OGE) process,
and the linear confinement term, besides a irrelevant constant\cite{TMNS01,MNSTU01I,BSS95}.
The linear potential at the long distance can be physically 
interpreted with the flux-tube picture or the string picture 
for hadrons\cite{N74,KS75,CNN79,CKP83,CI86,IP8385,SST95,SSTI95} with 
the string tension $\sigma \simeq $ 0.89 GeV/fm, which  
represents the magnitude of the confinement force. 
In this picture, the quark and the antiquark are linked with a 
one-dimensional flux-tube with the string tension $\sigma_{\rm Q\bar{Q}}$, 
and hence the Q-$\bar{\rm Q}$ potential is proportional to 
the distance $r$ between quark and antiquark at the long distance. 
This flux-tube picture (or the string picture) in the infrared region 
is supported by the Regge trajectory of hadrons\cite{GS94,SSSSG91}, 
the phenomenological analysis of heavy quarkonia data\cite{LSG91}, 
the strong-coupling expansion of QCD\cite{KS75}
and recent lattice QCD simulations\cite{GMO90,PH93,HSPW96}.

However, there is almost no reliable formula to describe the three-quark (3Q)
potential $V_{\rm 3Q}$ directly based on QCD, besides the strong-coupling QCD\cite{KS75,CI86}, 
although $V_{\rm 3Q}$ is directly responsible for the baryon properties\cite{CI86,BPV95,RS91} 
and is a primary quantity in the hadron physics. In fact, the 3Q potential has been treated 
phenomenologically or hypothetically more than 20 years.
In contrast with a number of studies on the Q-$\bar {\rm Q}$ 
potential using lattice QCD \cite{BS92,K00}, 
there were only a few lattice QCD studies for the 3Q potential  
done mainly more than 14 years ago \cite{SW8486,KEFLM87,TES88,B01}. 

Even at present, the arguments on the 3Q potential seem to be 
rather controversial. In Refs.\cite{SW8486,TES88,B01,AFT01,AFT02}, 
the 3Q potential seemed to be expressed by a sum of two-body potentials, 
which supports the $\Delta$-type flux tube picture \cite{C96}.
On the other hand, Ref.\cite{TMNS01,KEFLM87,TMNS99,SMNT00,TSMN01} seemed to support the Y-type 
flux-tube picture \cite{CI86,BPV95} rather than the $\Delta$-type one. 
These controversial results may be due to the difficulty of the 
accurate measurement of the 3Q ground-state potential in lattice QCD. 
For instance, in Refs.\cite{SW8486,TES88}, the authors did not 
use the smearing for ground-state enhancement, and therefore their results  
may include serious contamination from the excited-state component. 
In Refs.\cite{B01,AFT01,AFT02}, the author showed a preliminary result 
only on the equilateral-triangle case without the fit analysis. 

In this paper, for more than 300 different patterns of the 3Q system, 
we perform the accurate measurement of the static 3Q potential 
in SU(3) lattice QCD at $\beta$=5.7, 5.8 and 6.0, 
using the smearing method to remove the excited-state contaminations 
and to obtain the true ground-state potential.
The contents are organized as follows.
In Section 2, we make a brief theoretical consideration on the form 
of the inter-quark potential based on QCD.
In Section 3, we explain the method of the lattice QCD measurement of 
the 3Q potential, referring the importance of the smearing technique  
and its physical meaning.
In Section 4, we present the lattice QCD data of the 3Q potential,  
which are accurately measured from the smeared 3Q Wilson loop 
in a model-independent manner.  
In Section 5, we perform the fit analysis of the lattice data with the Y-ansatz.
In Section 6, we examine the various fit analyses with the $\Delta$-ansatz and
a more general ansatz including the Y and the $\Delta$ ans\"atze in some limits.
Section 7 is devoted to the summary and the concluding remarks.

\section{Theoretical consideration on the 3Q potential}
The lattice QCD data themselves are measured in a model-independent way,   
however, it is useful to make a theoretical consideration on the potential 
form in the static Q-$\bar{\rm Q}$ and 3Q systems with respect to QCD. 
Although QCD is one of the most difficult theories in 
the theoretical particle physics, 
there are two analytical methods based on QCD: 
one is the perturbative QCD, which is rather established to describe 
the short-distance behavior, and the other is the strong-coupling 
expansion, which is expected to reflect the long-distance behavior of QCD.  
At the short distance, perturbative QCD would work well  
according to the asymptotic freedom, and 
the static inter-quark potential can be 
described with the Coulomb-type potential 
as the one-gluon-exchange (OGE) result. 
On the other hand, at the long distance at the quenched level, 
the flux-tube picture with the string tension $\sigma$ 
is expected to be applicable from the argument of 
the strong-coupling expansion of QCD \cite{KS75,CI86,BPV95}, 
which indicates a linear-type confinement potential 
proportional to the total flux-tube length. 
Of course, it is nontrivial that these simple arguments on 
ultraviolet and infrared limits of QCD hold for 
the intermediate region as $0.2 \,{\rm fm} < r < 1 \,{\rm fm}$. 
Nevertheless, for instance, the lattice QCD data of 
the Q-$\bar {\rm Q}$ ground-state potential are well fitted by 
\begin{equation}
V_{\rm Q \bar{Q}}(r)=-\frac{A_{\rm Q \bar{Q}}}{r}
+\sigma_{\rm Q \bar{Q}} r+C_{\rm Q \bar{Q}}  
\end{equation}
at the quenched level \cite{BSS95}. 
Also in the phenomenological aspect of QCD, 
such a Q-${\bar{\rm Q}}$ potential is known to be successful to reproduce  
the empirical data of the mass spectra and the decay rates 
of various heavy quarkonia \cite{LSG91}. 
In fact, the Q-$\bar {\rm Q}$ potential $V_{\rm Q \bar{Q}}$ is 
well described by a sum of the short-distance 
OGE result and the long-distance flux-tube result. 

Also for the 3Q ground-state potential $V_{\rm 3Q}$, we basically adopt 
this picture as a theoretical frame of reference. 
In the 3Q system in the color-flux-tube picture, 
reflecting the character of SU($N_c$=3) in QCD, 
there can appears the physical junction linking 
to the three flux tubes stemming from the valence quarks.  
Since the confinement part is proportional to 
the total flux-tube length in this picture,   
the physical junction is expected to appear at the Fermat point 
of the 3Q triangle, as shown in Fig.~\ref{lmin}, as long as 
the ground-state 3Q system with spatially fixed valence quarks is concerned.  
Here, the Fermat point is defined so as to minimize the sum of 
the distances to the three vertices of the triangle. 

For the convenient description in the argument of the 
ground-state 3Q potential, 
we denote by $L_{\rm min}$ the minimal value of 
the total length of color flux tubes linking the three quarks.
When all angles of the 3Q triangle do not exceed $2\pi/3$, 
$L_{\rm min}$ is expressed as 
\begin{equation}
L_{\rm min}=\left[\frac12 (a^2+b^2+c^2)
 +\frac{\sqrt{3}}2 
  \sqrt{(a+b+c)(-a+b+c)(a-b+c)(a+b-c)}\right]^{\frac12}, 
\end{equation}
where $a$,$b$ and $c$ denote the three sides of the 3Q triangle 
as shown in Fig.~\ref{lmin}.
In this case, the physical junction appears and 
connects the three flux tubes originating from the three quarks, 
and the shape of the 3Q system is expressed as a Y-type flux 
tube \cite{KS75,CI86}, where the angle between two flux tubes is found 
to be $2\pi/3$ \cite{CI86,BPV95}. 
When an angle of the 3Q triangle exceeds $2\pi/3$, one finds 
\begin{equation}
L_{\rm min}=a+b+c-{\rm max}(a,b,c).
\end{equation}

In the picture of the short-distance OGE result 
plus the long-distance flux-tube result, 
the 3Q ground-state potential $V_{\rm 3Q}$ is expected to take a form of 
\begin{equation}
V_{\rm 3Q}=-A_{\rm 3Q}\sum_{i<j}\frac1{|{\bf r}_i-{\bf r}_j|}
+\sigma_{\rm 3Q} L_{\rm min}+C_{\rm 3Q}, 
\label{yansatz}
\end{equation}
which is referred to as the 
Y-ansatz~\cite{TMNS01,TSMN01,BPV95}. 
In the following sections, we will extract first the lattice QCD data of the 
3Q potential without any model assumption, and later we will try 
the fit analysis of the lattice data with the Y-ansatz or 
other possible ans\"atze. 

\section{The lattice QCD measurement for the 3Q potential} 
\subsection{The 3Q Wilson loop and the 3Q potential in QCD}

\begin{figure}[p]
\begin{center}
\epsfile{file=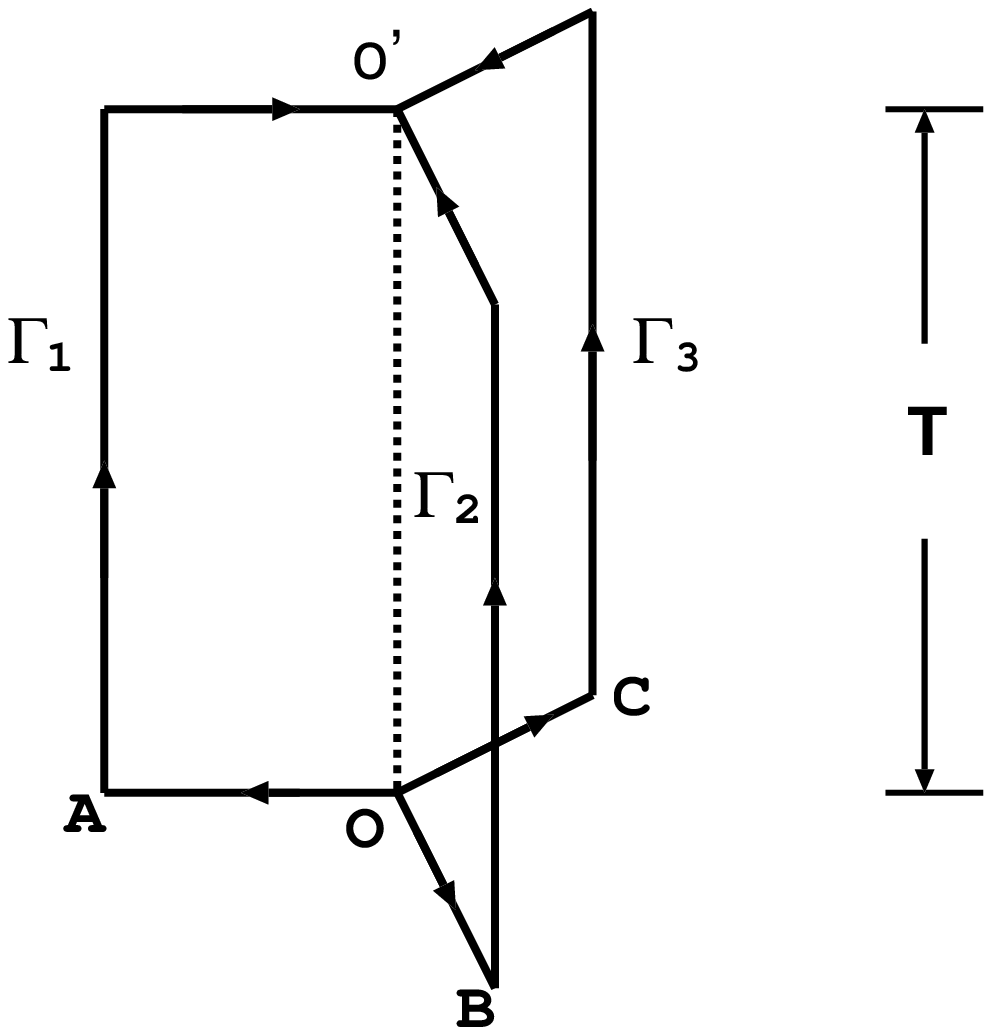,height=4cm}
\caption{The 3Q Wilson loop $W_{\rm 3Q}$. The 3Q state is generated at $t=0$ and is
annihilated at $t=T$. The three quarks are spatially fixed in ${\bf R}^3$ for $0 < t < T$.}
\label{bloop}
\end{center}
\end{figure}

Similar to the derivation of the Q-${\bar {\rm Q}}$ potential 
from the Wilson loop, the 3Q static potential $V_{\rm 3Q}$ 
is obtained with the 3Q Wilson loop as 
\begin{equation}
V_{\rm 3Q}=-\lim_{T \rightarrow \infty} \frac1T 
\ln \langle W_{\rm 3Q}\rangle.
\end{equation}
The 3Q Wilson loop $W_{\rm 3Q}$ is defined in a gauge-invariant manner as 
\begin{equation}
W_{\rm 3Q} \equiv \frac1{3!}\varepsilon_{abc}\varepsilon_{a'b'c'}
U_1^{aa'} U_2^{bb'} U_3^{cc'} 
\label{3qope}
\end{equation}
with the path-ordered product 
\begin{equation}
U_k \equiv P \exp\{ig\int_{\Gamma_k}dx_\mu A^{\mu}(x)\} 
(k=1,2,3),  
\end{equation}
along the path denoted by ${\Gamma_k}$ in Fig.~\ref{bloop}. 
As shown in Fig.~\ref{bloop}, the 3Q Wilson loop physically expresses 
the 3Q gauge-invariant state which 
is generated at $t=0$ and is annihilated at $t=T$ 
with the three quarks spatially fixed in ${\bf R}^3$ for $0 < t < T$.

The initial (or the finial) 3Q state is introduced as 
the string-like object in the naive 3Q Wilson loop. 
However, the physical ground state of the 3Q system, 
which is of interest here, 
is expected to be expressed by the flux tubes instead of the strings, 
and then the 3Q state which is expressed by the strings 
generally includes excited-state components such 
as flux-tube vibrational modes. 
Of course, if the large $T$ limit can be taken, the 
ground-state potential would be obtained. However, 
the practical measurement of $\langle W_{\rm 3Q}\rangle$ 
is rather severe for large $T$ in lattice QCD calculations, because 
$\langle W_{\rm 3Q}\rangle$ decreases exponentially with $T$.

Therefore, for the accurate measurement of the 3Q ground-state 
potential $V_{\rm 3Q}$, it is practically indispensable 
to reduce the excited-state components in the 3Q system 
introduced at $t=0$ and $t=T$ in the 3Q Wilson loop. 
The gauge-covariant smearing method is one of the most useful technique 
for the ground-state enhancement \cite{SMNT00,APE87,BSS95} 
without breaking the gauge covariance, and  
is adopted to measure the Q-$\bar {\rm Q}$ potential 
and the glueball mass \cite{ISM02} in the recent lattice QCD calculation. 
(This smearing method was not applied to a few pioneering 
lattice studies on the 3Q potential\cite{SW8486,TES88}, 
since the smearing technique was mainly developed after their works.
As will be discussed later, their numerical results 
seem to include fatal large excited-state contaminations.)

In this paper, we perform the accurate measurement of 
the 3Q ground-state potential $V_{\rm 3Q}$, 
using the ground-state enhancement by the gauge-covariant
smearing method for the link-variable
in SU(3)$_c$ lattice QCD at the quenched level \cite{SMNT00}.

\subsection{The smearing method for the ground-state enhancement}

Let us consider here the physical states of the 3Q system with 
the spatially fixed quarks. 
In this 3Q system, of course, there is no valence-quark motion, and 
the central issue is the gluonic configuration 
under the boundary condition of the spatially fixed three quarks, 
which play the role of the color source of the gluonic color-electric flux. 

Like the Q-$\bar{\rm Q}$ flux-tube system, 
the ground state of the 3Q system is expected to be composed 
by flux tubes rather than the strings\cite{TMNS01,SST95}, and 
there are many excited states of the 3Q system 
corresponding to the flux-tube vibrational modes\cite{TMNS01}. 
We here express the 3Q Wilson loop with the normalized physical states, 
the 3Q ground state $|{\rm g.s.;} t \rangle$ and 
the $k$-th excited 3Q state $|k {\rm th \ e.s.;} t \rangle$ at $t$. 
In the 3Q Wilson loop, the normalized gauge-invariant 3Q state 
$|{\rm 3Q}; 0\rangle$ created at $t=0$ and $|{\rm 3Q}; T\rangle$ 
annihilated at $t=T$ can be expressed as
\begin{eqnarray}
&&|{\rm 3Q}; 0\rangle =c_0|{\rm g.s.;0}\rangle+c_1|{\rm 1st \ e.s.;0}\rangle+
c_2|{\rm 2nd \ e.s.;0}\rangle+...,\\ \nonumber
&&|{\rm 3Q}; T\rangle =c_0|{\rm g.s.;} T \rangle+c_1|{\rm 1st \ e.s.;} T \rangle+
c_2|{\rm 2nd \ e.s.;} T \rangle+...,
\end{eqnarray}
with the coefficients $c_i$ obeying the normalization condition 
$\sum_{i=0}^\infty |c_i|^2=1$. 
Then, the expectation value of $W_{\rm 3Q}$ can be expressed as
\begin{eqnarray}
\langle W_{\rm 3Q}(T)\rangle&=& \langle {\rm 3Q}; T|{\rm 3Q}; 0 \rangle= 
|c_0|^2\langle{\rm g.s.} ; T|{\rm g.s.;0}\rangle+
|c_1|^2\langle{\rm 1st \ e.s.;}T|{\rm 1st \ e.s.;0}\rangle+...\\ \nonumber
&=& |c_0|^2\exp(-V_{\rm g.s.}T)+|c_1|^2\exp(-V_{\rm 1st\ e.s.}T)+...,
\label{expan}
\end{eqnarray}
with the ground-state potential $V_{\rm g.s.}$ and 
the $k$-th excited-state potential $V_{k\hbox{-}{\rm th\ e.s.}}$, which 
correspond to the energy-eigenvalues of the 3Q system.
(Note that the normalization here is consistent with 
the definition of $W_{\rm 3Q}$ in Eq.(\ref{3qope}), which leads to 
$\langle W_{\rm 3Q}(T=0) \rangle$ =1.) 

As increasing $T$, the excited-state components drop faster than the 
ground-state component in $\langle W_{\rm 3Q} \rangle$, however, the 
ground-state component $|c_0|^2\exp(-V_{\rm g.s.}T)$ also 
decreases exponentially. 
Hence, we face a practical difficulty in extracting the numerical signal.
To avoid this difficulty, we adopt the smearing 
technique\cite{TMNS01,BSS95,APE87} which enhances the ground-state 
overlap as $|c_0|^2$ and removes the excited-state contamination efficiently.

\begin{figure}[p]
\begin{center}
\epsfile{file=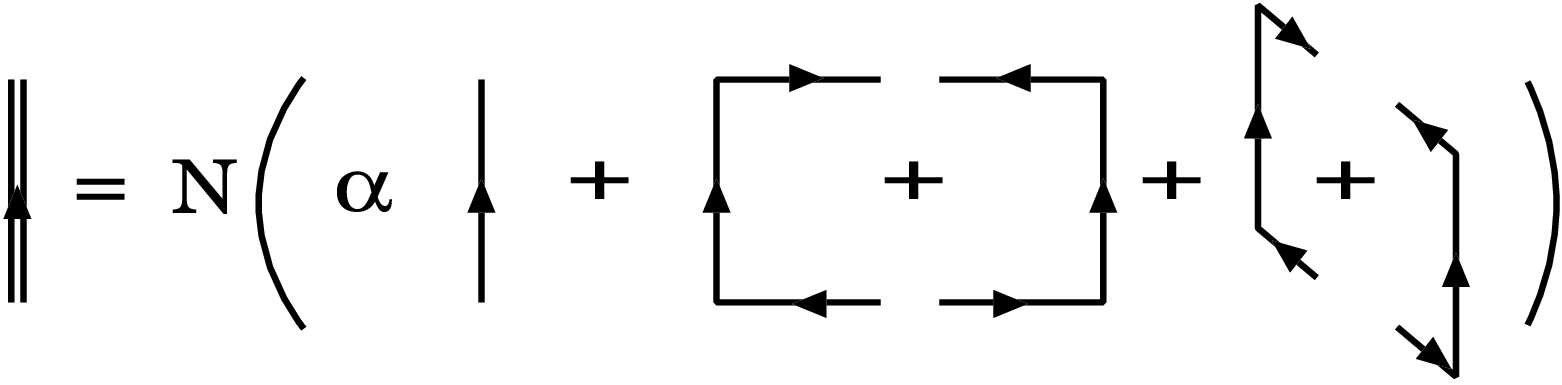,height=2.5cm}
\caption{The schematic explanation of the smearing for the link-variables.}
\label{smearing1}
\end{center}
\end{figure}

The smearing method is one of the most popular and useful 
techniques to extract the ground-state potential in lattice QCD. 
The standard smearing for link-variables is expressed as the 
iterative replacement of the spatial link-variable $U_i (s)$ 
($i=1,2,3$) by the obscured link-variable 
$\bar U_i (s) \in {\rm SU(3)}_c$ \cite{BSS95,APE87} which maximizes  
\begin{equation}
{\rm Re} \,\,{\rm tr} \{ \bar U_i(s) V_i^{\dagger}(s) \}
\end{equation}
with 
\begin{equation}
V_i(s) \equiv \alpha U_i(s)+\sum_{j \ne i} \{
U_j(s)U_i(s+\hat j)U_j^\dagger (s+\hat i) + 
U_j^\dagger (s-\hat j)U_i(s-\hat j)U_j(s+\hat i-\hat j)\}, 
\end{equation}
which is schematically illustrated in Fig.~\ref{smearing1}.
Here, $\alpha \in {\bf R}$ is referred to as the smearing parameter.  
The $n$-th smeared link-variables $U_\mu^{(n)}(s)$ $(n=1,2,..,N_{\rm smear})$ 
are iteratively defined starting from $U_\mu^{(0)}(s) \equiv U_\mu(s)$ as
\begin{equation}
U_i^{(n)}(s) \equiv \bar U_i^{(n-1)}(s) 
\quad 
(i=1,2,3), 
\qquad 
U_4^{(n)}(s) \equiv U_4(s).
\end{equation}

For arbitrary operator $F[U_\mu(\cdot)]$, 
the $n$-th smeared operator $F[U_\mu^{(n)}(\cdot)]$ is defined 
with the $n$-th smeared link-variable $U_\mu^{(n)}(s)$ instead of the 
original link-variable.
The inter-quark potential can be accurately measured from the 
properly smeared (3Q) Wilson loop. 
Here, the smearing parameter $\alpha$ and the iteration number $n$ 
play the role of the variational parameters and 
are properly chosen so as to maximize the ground-state component.

We note that the smearing is just a method to 
choose the flux-tube-like operator, and hence 
it never changes the physics itself such as the gauge configuration. 
As an important feature, this smearing procedure keeps 
the gauge covariance of the ``fat'' link-variable $U_\mu^{(n)}(s)$ properly. 
In fact, the gauge-transformation property of 
$U_\mu^{(n)}(s)$ is just the same as that of the original link-variable 
$U_\mu(s)$, and therefore 
the gauge invariance of $F(U_\mu^{(n)}(s))$ is ensured 
for the arbitrary gauge-invariant operator $F(U_\mu (s))$. 
For instance, the $n$-th smeared (3Q) Wilson loop is gauge-invariant. 

While no temporal extension appears in the smearing, 
the fat link-variable $U_\mu^{(n)}(s)$ includes a spatial extension 
in terms of the original link-variable $U_\mu(s)$, 
and then the smeared ``line'' expressed with $U_\mu^{(n)}(s)$ physically 
corresponds to a ``flux tube'' with the spatial extension.
Therefore, if a suitable smearing is done, the smeared line 
is expected to be close to the ground-state flux tube. 
This smearing method is actually successful for the extraction 
of the Q-$\bar {\rm Q}$ potential in lattice QCD \cite{BSS95}.

\subsection{The physical meaning of the smearing method}
We consider here the physical meaning of the 
smearing method with the smearing parameter $\alpha$ 
in terms of the size or the spatial extension of the $n$-th smeared line.  
For the convenience of the description, we define
\begin{equation}
	p    \equiv   {\alpha \over \alpha+4}, \qquad
	q \equiv {1 \over \alpha+4},
\end{equation}
which  satisfy   $p+4q=1$. 
Let us consider the smearing of the line-like object, 
which is idealized to be infinitely long. 
As mentioned above, the smeared line corresponds to the 
spatially extended flux tube in terms of the original link-variable. 
Here, we locate the $n$-th smeared line on the $z$-axis in ${\bf R}^3$, 
and then, due to the translational invariance along the $z$-direction, 
the argument is essentially two-dimensional and depends only on $x$ and $y$, 
and the flux direction is to be in the $z$-direction. 
We denote by $\varphi(x,y;n)$ ($x, y \in {\bf R}$) 
the spatial flux distribution in the $n$-th smeared line. 
(On the lattice with the spacing $a$, 
$\varphi(x,y;n)$ is defined on the discrete 
points $(x,y)=(n_x a, n_y a)$ with $n_x,n_y \in {\bf Z}$.)

\begin{figure}[p]
\begin{center}
\epsfile{file=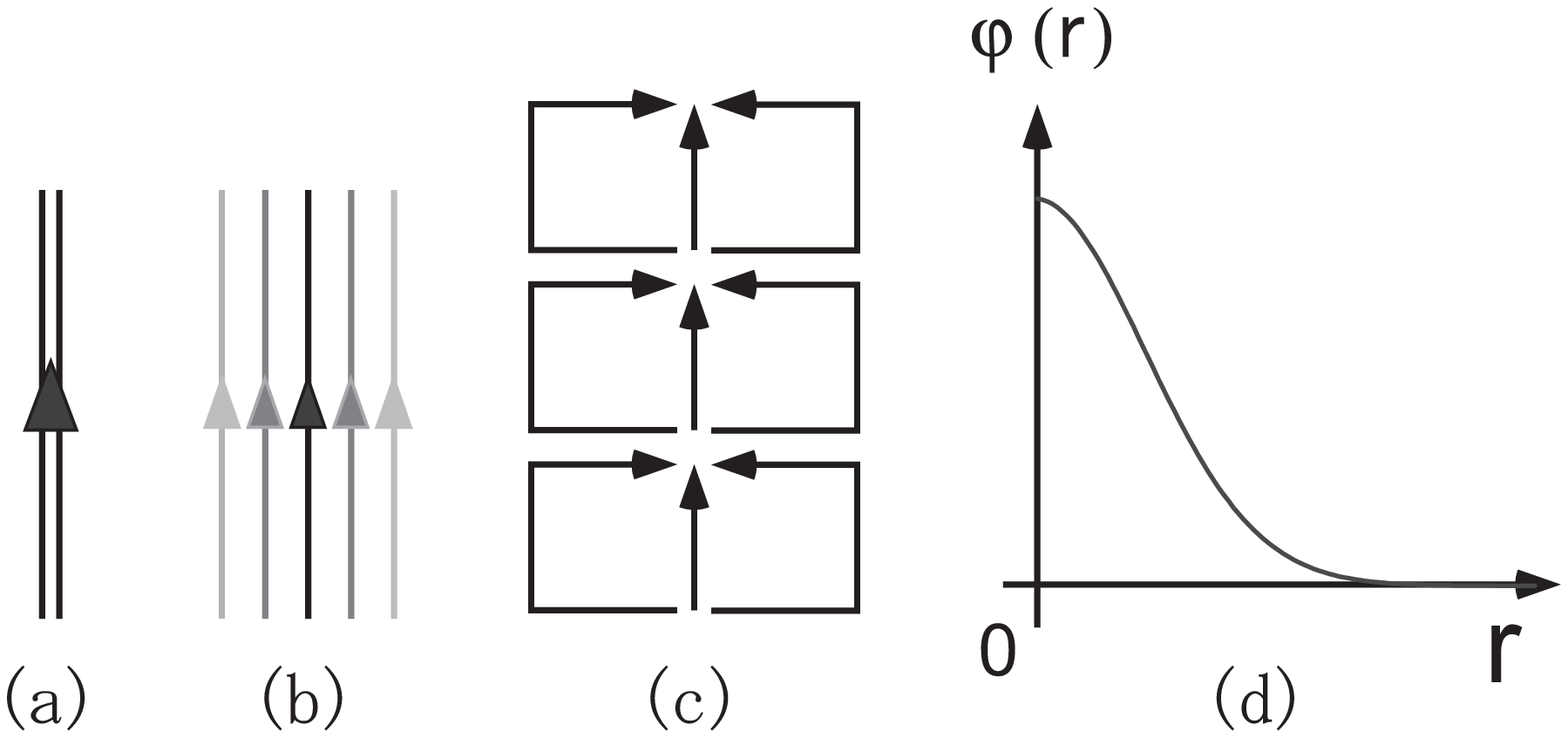,height=6cm}
\caption{
The schematic explanation of the physical meaning of the smeared line.
The $n$-th smeared line depicted as (a) physically corresponds to the spatially-distributed
flux tube as (b) in terms of the original field variable.
The single smearing procedure for the line is illustrated with (c) on the lattice.
The flux perpendicular to the line is expected to be canceled.
The spatial distribution of the $n$-th smeared line is expressed by the Gaussian profile
$\phi(r)$ with $r=(x^2+y^2)^{1/2}$ as shown in (d).}
\label{smearing2}
\end{center}
\end{figure}

 From the iterative definition of the smearing,  
the spatial flux distribution $\varphi(x,y;n+1)$ of 
the \mbox{$(n+1)$-th} smeared line is expected to relate to  
$\varphi(x,y;n)$ of the $n$-th smeared line as  
\begin{eqnarray}
	\varphi(x,y;n+1)=p \varphi(x,y;n)+
	q\{\varphi(x+a,y;n)+\varphi(x-a,y;n)
          +\varphi(x,y+a;n)+\varphi(x,y-a;n)\}.
\end{eqnarray}
Here, as shown in Fig.~\ref{smearing2}~(c), we assume the cancellation of  
non-$z$-components of the flux, which exactly holds for the abelian flux. 
Using the difference operator, we obtain  
\begin{eqnarray}
\Delta_n \varphi(x,y;n) 
&\equiv& \varphi(x,y;n+1) - \varphi(x,y;n) \\ \nonumber
&=& q\{\Delta_x \varphi(x,y;n)-\Delta_x \varphi(x-a,y;n)
+\Delta_y \varphi(x,y;n)-\Delta_y \varphi(x,y-a;n) \} \\ \nonumber
&=& q\{\Delta_x^B \Delta_x \varphi(x,y;n)
      +\Delta_y^B \Delta_y \varphi(x,y;n)\}, 
\end{eqnarray}
where $\Delta_k$ and $\Delta_k^B$ ($k=x,y$) denote 
the forward and the backward difference operators satisfying 
$\Delta_k f(\vec r) \equiv f(\vec r+\hat k)- f(\vec r)$ and   
$\Delta^B_k f(\vec r) \equiv f(\vec r)- f(\vec r-\hat k)$, 
respectively. 

When the lattice spacing $a$ is small enough, 
the spatial difference $\Delta_k$ can be approximated 
by the spatial derivative as $\Delta_k \simeq a \partial_k$. 
Also for the iteration number $n$ of the smearing, we formally introduce a 
small ``spacing'' $a_n$  in the ``$n$-direction'', 
and define the semi-continuum parameter $\tilde n \equiv n a_n$, 
although the final result does not depend on the artificial spacing $a_n$.
Then, the difference $\Delta_n$ can be approximated by the derivative 
as $\Delta_n \simeq a_n \partial_{\tilde n}$. 
In this way, we obtain the differential equation as 
\begin{equation}
	{\partial \over \partial \tilde n}
	\varphi(x,y;n)= D (\partial_x^2+\partial_y^2) \varphi(x,y;n),
\end{equation}
which corresponds to the ``diffusion equation'' at the ``time'' $\tilde n$ 
with the diffusion parameter 
\begin{equation}
D \equiv \frac{q a^2}{a_n}
=\frac1{\alpha +  4}\frac{a^2}{a_n}.
\end{equation}
The ``initial condition'' at $n=0$ is given as 
\begin{equation}
\varphi(x,y;n=0)=\delta(x)\delta(y),
\end{equation}
which means the simple line before applying the smearing.
Then, the flux distribution $\varphi(x,y;n)$ in the $n$-th smeared line 
can be expressed as 
\begin{equation}
	\varphi(x,y;n)=
	{1\over (4\pi D \tilde n)}
	\exp\left[ - {x^2 + y^2 \over 4 D \tilde n} \right].
\end{equation}
Thus, the $n$-th smeared line  
physically corresponds to the Gaussian spatially-distributed flux tube   
in terms of the original link-variable as shown in 
Fig.~\ref{smearing2}~(d). 

As a result, the flux-tube size can be roughly estimated as
\begin{equation}
R \equiv \sqrt{\langle x^2+y^2 \rangle}
\equiv \left(\frac{\int dx dy \varphi(x,y;n) (x^2+y^2)}
{\int dx dy \varphi(x,y;n)}\right)^{1/2}
=
	2\sqrt{D \tilde n}
=
	2a \sqrt{n \over \alpha + 4}.
\label{size}
\end{equation}
We note that the square root appears 
as a character of the Brownian motion, 
and hence $n$-dependence of the flux-tube radius $R$ is not so strong. 
This formula also explains the physical roles of the 
two parameters, $\alpha$ and $n$. 
The smearing  parameter $\alpha$ controls the speed of the smearing, 
and the speed of smearing is slower for larger $\alpha$. 
For each fixed $\alpha$, $n$ plays the role
of extending the size of the smeared operator.  
Hence, once we find the suitable smearing parameters $n$ and
$\alpha$ which achieve a large ground-state overlap, the physical 
size of the flux tube is roughly estimated with Eq.(\ref{size}).

\section{The lattice QCD results of the 3Q potential}
We measure the 3Q potential from the properly smeared 3Q Wilson loop 
in SU(3)$_c$ lattice QCD at the quenched level. In this section, 
we present the lattice QCD data of the 3Q ground-state potential $V_{\rm 3Q}$ 
for more than 300 different patterns of the 3Q systems in total. 
These lattice QCD data are, of course, the model-independent data based on QCD, 
and we think that the data themselves are useful for the study of the 3Q system, 
particularly for the phenomenological approach as the quark model for baryons.

\subsection{The simulation conditions of lattice QCD}
The gauge configurations are generated using the SU(3)$_c$ lattice QCD 
Monte-Carlo simulation with the standard action with 
$12^{3} \times 24$ at $\beta=5.7$ and 
$16^{3} \times 32$ at $\beta=5.8,6.0$ at the quenched level.
The pseudo-heat-bath algorithm is adopted for update of the gauge configuration.
After a thermalization of more than 5,000 sweeps, 
we sample the gauge configuration every 500 sweeps, and 
we use at least 150 gauge configurations at each $\beta$ 
for the study of the 3Q potential.
We summarize in Table~\ref{gaugeprm}
the lattice parameters and the related information 
on the simulation as well as the lattice spacing $a$ determined 
so as to reproduce the string tension as $\sigma$=0.89 GeV/fm
in the Q-$\bar{\rm Q}$ potential $V_{\rm Q \bar{Q}}$ at each $\beta$.
As for the smearing, we set the smearing parameter as $\alpha$=2.3, which is one of the most 
suitable smearing parameter for the calculation of the 3Q ground-state potential.
The iteration number $N_{\rm smr}$ of the smearing which maximizes the ground-state overlap 
and the corresponding flux-tube radius $R$ are also listed at each $\beta$ 
in Table~\ref{gaugeprm} . 
On the statistical error of the lattice data, we adopt the 
jackknife error estimate \cite{MM94}.
The Monte-Carlo simulations at $\beta=5.7, 6.0$ and 5.8 have been performed 
on NEC-SX4, NEC-SX5 at Osaka University and HITACHI-SR8000 at KEK, respectively.

\subsection{The Q-$\bar {\bf Q}$ potential}

\begin{figure}[p]
\begin{center}
\epsfile{file=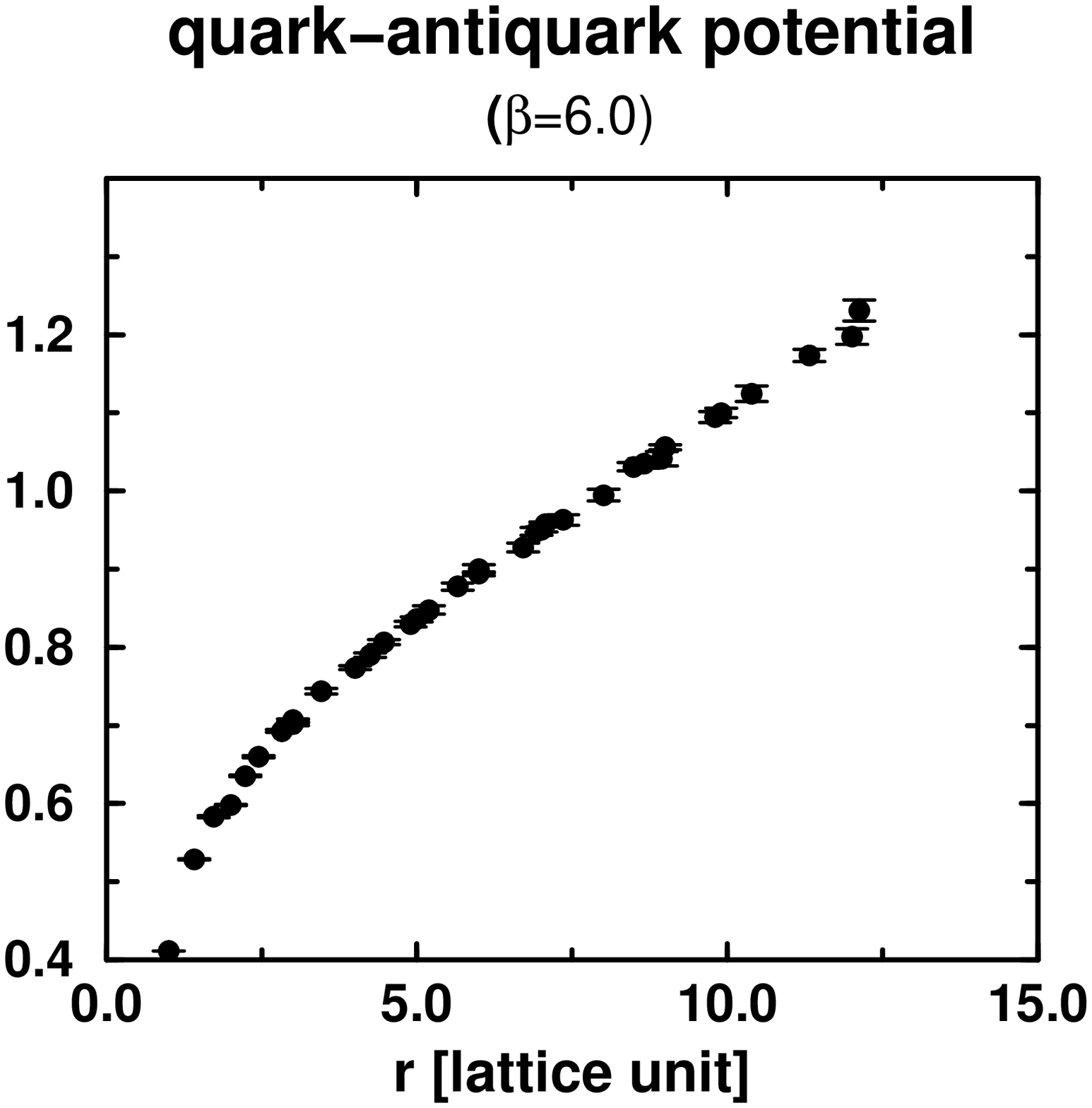,height=7cm}
\caption{
The ${\rm Q\bar{Q}}$ static potential $V_{\rm Q\bar{Q}}(r)$
as the function of the inter-quark distance $r$ in the lattice unit
in SU(3) lattice QCD with $\beta=6.0$ at the quenched level.
}
\label{qqbar60}
\end{center}
\end{figure}

As a frame of reference, we measure the Q-$\bar{\rm Q}$ potential $V_{\rm Q \bar{Q}}$ 
from the properly smeared Wilson loop in the present lattice QCD. 
As is consistent with the previous lattice works\cite{BSS95}, the lattice QCD data of 
the Q-$\bar {\rm Q}$ ground-state potential $V_{\rm Q \bar{Q}}$ at the quenched level 
are well reproduced by 
\begin{equation}
V_{\rm Q \bar{Q}}(r)=-\frac{A_{\rm Q \bar{Q}}}{r}
+\sigma_{\rm Q \bar{Q}} r+C_{\rm Q \bar{Q}}  
\label{qqbarfit}
\end{equation}
with the best-fit parameter set   
($A_{\rm Q \bar{Q}}$, $\sigma_{\rm Q \bar{Q}}$, $C_{\rm Q \bar{Q}}$) listed in 
Table~\ref{fitprm}  at each $\beta$. As a visual illustration, we show in 
Fig.~\ref{qqbar60} the lattice QCD data of $V_{\rm Q \bar{Q}}(r)$ 
as the function of the inter-quark distance $r$ in the lattice unit. 
One finds a good agreement of the lattice data of $V_{\rm Q \bar{Q}}$ 
and the fit curve with Eq.(\ref{qqbarfit}).

In spite of the visual agreement, the statistical analysis is also necessary for the 
argument on the fit. This is rather difficult because the lattice QCD data include 
not only the statistical error but also the systematic errors from the discretization, 
which cannot be estimated straightforwardly.
We here examine the on-axis and the off-axis data at each $\beta$. 
The on-axis data of $V_{\rm Q \bar{Q}}$ are well fitted with Eq.(\ref{qqbarfit}) at 
each $\beta$. However, when we include the off-axis data, the fit of $V_{\rm Q \bar{Q}}$ 
with Eq.(\ref{qqbarfit}) becomes rather worse as $\chi^2/N_{\rm DF} \sim 10$. 
This is due to the breaking of the rotational invariance on the lattice, and 
such breaking is significant for the short-distance data.

As will be discussed in Section 5-5, to be strict, the lattice Coulomb potential would be 
preferable instead of the Coulomb potential at least for the short-distance lattice data.
We examine also the fit with the lattice Coulomb plus linear potential, 
and list the best-fit parameter set ($A_{\rm Q \bar{Q}}^{\rm LC}$, 
$\sigma_{\rm Q \bar{Q}}^{\rm LC}$, $C_{\rm Q \bar{Q}}^{\rm LC}$) at each $\beta$ 
together with $\chi^2/N_{\rm DF}$ in Table~\ref{fitprm}, where the label as ``off-axis'' 
means the fit analysis for both on-axis and off-axis data. We then find that the fit with 
the lattice Coulomb plus linear potential for both on-axis and off-axis data is fairly good, 
although the fit parameters such as the string tension are almost unchanged.

\subsection{The ground-state enhancement through the smearing}

Before presenting the lattice data of the 3Q potential, we briefly demonstrate the utility 
of the smearing method by estimating the magnitude of the ground-state component in  
the 3Q state at $t=0, T$ in the smeared 3Q Wilson loop $W_{\rm 3Q}$, which is 
composed with the $n$-th smeared link-variable $U_\mu^{(n)}(s)$. 

 From the similar argument in the Q-$\bar{\rm Q}$ system \cite{BSS95}, 
the overlap of the 3Q-state operator with the ground state is estimated with 
\begin{equation}
C_0 \equiv \frac{\langle W_{\rm 3Q} (T) \rangle ^{T+1}}
{\langle W_{\rm 3Q} (T+1) \rangle ^T}, 
\end{equation}
which is referred to as the ground-state overlap. For instance, in the ideal case where 
the 3Q state is the perfect ground state in the smeared 3Q Wilson loop, 
one gets $\langle W_{\rm 3Q} (T) \rangle = e^{-V_{\rm 3Q} T}$ and then $C_0$=1. 
(Here, $W_{\rm 3Q} (T)$ is normalized as $\langle W_{\rm 3Q} (T=0) \rangle=1$, 
as shown in Eq.(\ref{3qope}). In accordance with the excited-state contamination, 
$C_0$ is reduced to be a small value less than unity.

\begin{figure}[p]
\begin{center}
\epsfile{file=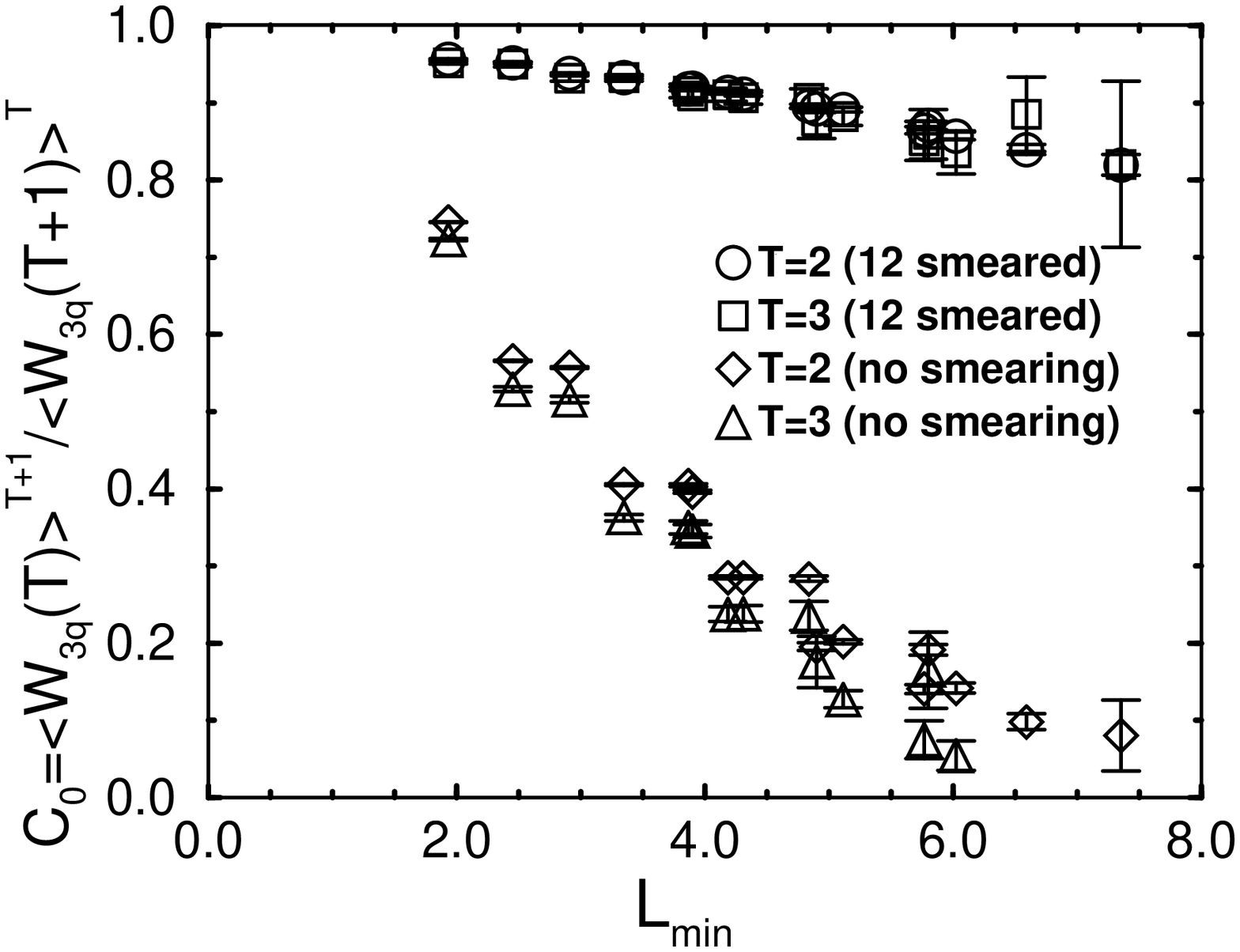,height=7cm}
\caption{
The ground-state overlap of the 3Q system,
$C_0 \equiv \langle W_{\rm 3Q} (T) \rangle ^{T+1} / \langle W_{\rm 3Q} (T+1) \rangle ^T$,
with the smeared link-variable (upper data) and with unsmeared link-variable (lower data)
at $\beta=5.7$.
To distinguish the 3Q system, we have taken the horizontal axis as $L_{\rm min}$,
which denotes the minimal value of the total length of the flux tubes linking the three quarks.
For each 3Q configuration, $C_0$ is largely enhanced as $0.8 < C_0 < 1$ by the smearing.
}
\label{czero}
\end{center}
\end{figure}

We note that the ground-state potential $V_{\rm 3Q}$ can be measured accurately 
if $C_0$ is large enough and is close to unity. 
Then, we check the ground-state overlap $C_0$ in the $n$-th smeared 3Q Wilson loop 
$\langle W_{\rm 3Q} (U_\mu^{(\cdot)}(s), T) \rangle$ using lattice QCD simulations, 
and search reasonable values of the smearing parameter $\alpha$ and the iteration number 
$N_{\rm smr}$ of the smearing so as to make $C_0$ large. 
For instance, the ground-state overlap $C_0$ is largely enhanced as $0.8 < C_0 < 1$ 
even for $T \le 3$ by the smearing with $\alpha=2.3$ and $N_{\rm smr}=12$ 
for all of the 3Q configurations at $\beta=5.7$ as shown in Fig.~\ref{czero}. 
Thus, the ground-state component is largely enhanced by the suitable smearing. 

We list in Table~\ref{gaugeprm} one of the best parameter set ($\alpha, N_{\rm smr}$) 
at each $\beta$. We note that, as will be shown in the next subsection, 
the magnitude of the ground-state overlap can be also estimated with 
$\bar C$ in Table~\ref{573qpot}-\ref{603qpot4}. One finds a large value of 
$\bar C$ close to unity as $\bar C \ge 0.7$ for each lattice data on the 3Q system.

\subsection{The lattice QCD data of the 3Q potential}

Now, we perform the accurate measurement of the 3Q ground-state potential $V_{\rm 3Q}$ 
using the smearing technique in SU(3) lattice QCD.
We investigate more than 300 different patterns of the 3Q systems in total.
In the practical calculation, we consider the following two type 3Q system on the lattice.

\begin{enumerate}
\renewcommand{\labelenumi}{(\Roman{enumi})}
\item The 3Q system where the three quarks are put on the three spatial axes as 
$(i,0,0)$, $(0,j,0)$, $(0,0,k)$ ($i,j,k=0,1,2,...$) in ${\bf R}^3$ in the lattice unit. 
\item The 3Q system where the three quarks are put on the $x$-$y$ plane as 
$(l,0,0)$, $(-m,0,0)$, $(0,n,0)$ ($l,m,n=0,1,2,...$) in ${\bf R}^3$ in the lattice unit.
\end{enumerate}

In both cases, the junction point O in the 3Q Wilson loop is 
set at the origin $(0,0,0)$ in ${\bf R}^3$, 
although the final result of the ground-state potential 
$V_{\rm 3Q}$ should not depend on the artificial selection of O. 
(As will be shown in the fit analysis in Section 5, there is no discontinuity 
between (I) and (II) on the 3Q potential $V_{\rm 3Q}$, in spite of the fairly different 
setting of the artificial junction O. This suggests that $V_{\rm 3Q}$ is independent of O.)
For each pattern of the 3Q system, 
we calculate the 3Q Wilson loop of all equivalent 3Q systems 
by changing O and the direction of $\hat{x},\hat{y},\hat{z}$, 
using the translational, the rotational and the reflection symmetries on lattices. 

Owing to the smearing, the ground-state component is largely enhanced,  
and therefore the 3Q Wilson loop $\langle W_{\rm 3Q} \rangle$ 
composed with the smeared link-variable exhibits a single-exponential behavior as 
\begin{equation}
\langle W_{\rm 3Q} \rangle \simeq e^{-V_{\rm 3Q}T}
\end{equation}
even for a small value of $T$. 

For each 3Q configuration, we measure 
$V_{\rm 3Q}^{\rm latt}$ from the least squares fit with the single-exponential form
\begin{equation}
\langle W_{\rm 3Q}\rangle =\bar{C}e^{-V_{\rm 3Q}T}.
\label{extv3q}
\end{equation}
Here, we choose the fit range of $T$ such that the stability of the ``effective mass''
\begin{equation}
V(T)\equiv \ln \frac{\langle W_{\rm 3Q}(T) \rangle}{\langle W_{\rm 3Q}(T+1)\rangle}
\end{equation}
is observed to avoid the effect of the excited-state contamination remaining at the small $T$ region.
In fact, we use a relatively large value on $T$ as the fit range for the accurate 
measurement.

In Table~\ref{573qpot}, we list the lattice QCD data $V_{\rm 3Q}^{\rm latt}$ 
of the 3Q ground-state potential at $\beta=5.7$, 
together with the prefactor $\bar{C}$ in Eq.(\ref{extv3q}), 
the fit range of $T \in [T_{\rm min}, T_{\rm max}]$ and $\chi^2/N_{\rm DF}$. 
In Table~\ref{583qpot1}-\ref{603qpot4},
we list up the lattice QCD data $V_{\rm 3Q}^{\rm latt}$ 
of the 3Q ground-state potential at $\beta=5.8, 6.0$, 
together with the prefactor $\bar{C}$ in Eq.(\ref{extv3q}). 
The statistical error of $V_{\rm 3Q}^{\rm latt}$ is estimated with the jackknife method. 
We stress again that these lattice QCD data are the model-independent data based on QCD, 
and we think that the data themselves are useful for the study of the 3Q system, 
particularly for the phenomenological approach as the quark model for baryons.

We note that the prefactor $\bar C$ physically means the magnitude of the 
ground-state overlap in the smeared 3Q Wilson loop. 
In fact, the pure ground-state 3Q system leads to $\bar C=1$, and $1-\bar C$ corresponds to 
the contribution of the excited-state contamination. 
We find a large ground-state overlap as $\bar{C} \ge 0.7$ for all 3Q configurations.

 From the best smearing parameters, $\alpha=2.3$ and $N_{\rm smr}$, the flux-tube radius 
$R$ can be roughly estimated with Eq.(\ref{size}) at each $\beta$.
We then get a rough estimate of the flux-tube radius as 
$R \simeq 0.52\ {\rm fm}$ both at $\beta$=5.7, 5.8 and 6.0. 
This flux-tube radius $R$ seems consistent with the typical hadron size, and 
it cannot be negligible in comparison with the flux-tube length between 
the junction and the quark in the 3Q systems in consideration.
In fact, the 3Q systems listed in Table~\ref{573qpot}-\ref{603qpot4} are to be 
regarded as a flux-tube rather than the string-like object, and hence 
it is nontrivial whether the strong-coupling QCD can be applicable or not in such 3Q systems.
Nevertheless, the Y-ansatz from the simple string picture is found 
to work well for the lattice QCD data of the three-quark potential.

\section{The fit analysis of the 3Q potential with the Y-ansatz}
For the study of the 3Q potential $V_{\rm 3Q}$, we are interested in its large-distance 
behavior relating to the confinement force rather than the short-distance one.
The short-distance behavior of $V_{\rm 3Q}$ is expected to be described by     
the two-body Coulomb potential as the one-gluon-exchange (OGE) result in perturbative 
QCD, although it is nontrivial whether perturbative QCD works well 
at the intermediate distance as $r \sim 0.5 \ {\rm fm}$. 
The OGE result indicates also a simple relation on the Coulomb coefficients in 
the Q-$\bar {\rm Q}$ and the 3Q potentials as $A_{\rm 3Q} \simeq \frac12 A_{\rm Q \bar {\rm Q}}$.

\subsection{The long-distance behavior of the 3Q potential}
\begin{figure}[p]
\begin{center}
\epsfile{file=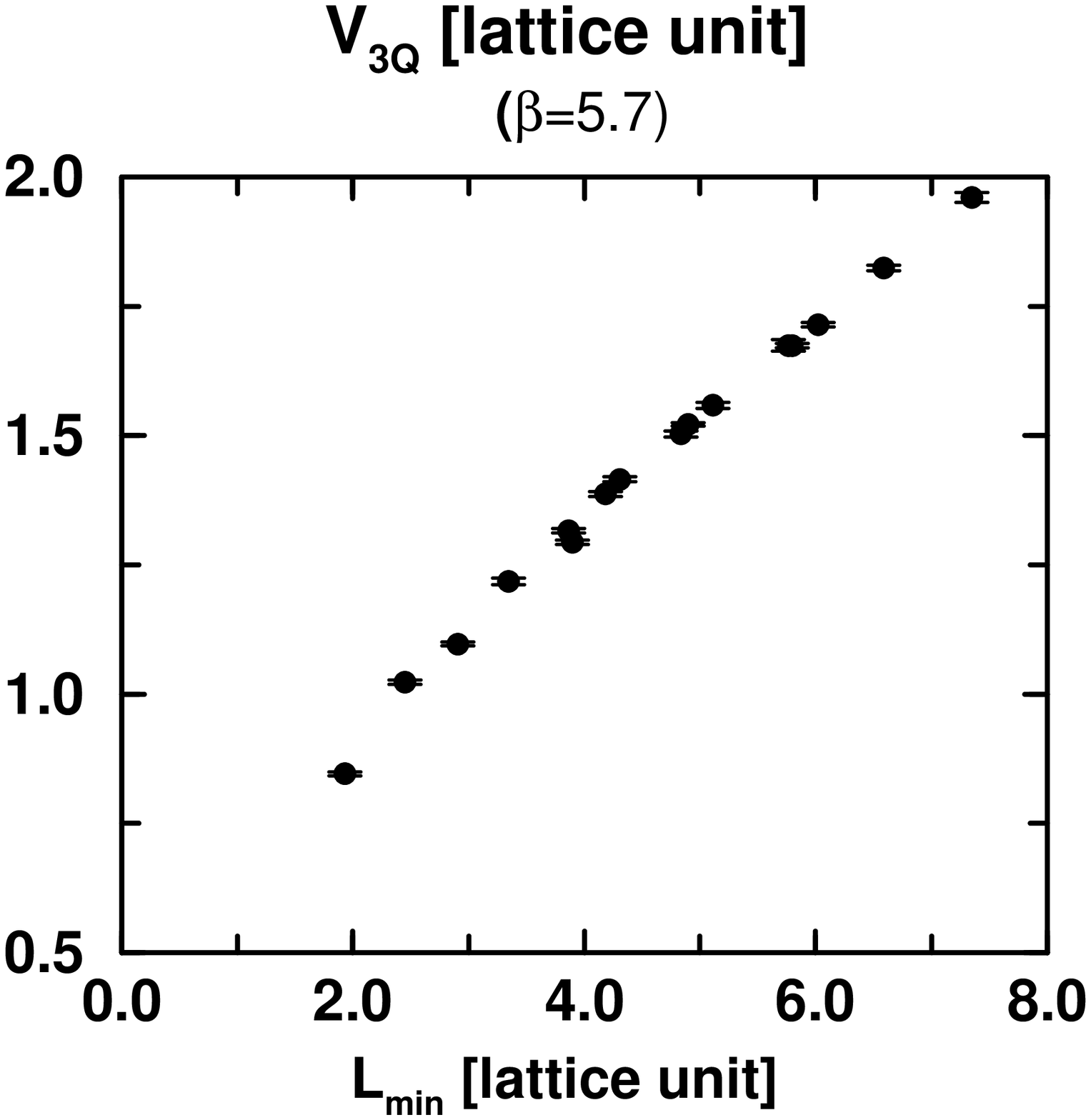,height=7cm}
\caption{
The lattice QCD data for the 3Q potential $V_{\rm 3Q}^{\rm latt}$ at $\beta=5.7$
as the function of $L_{\rm min}$, the minimum value of
the total length of the flux tubes, in the lattice unit.
}
\label{3q57nor}
\end{center}
\end{figure}

\begin{figure}[p]
\begin{center}
\epsfile{file=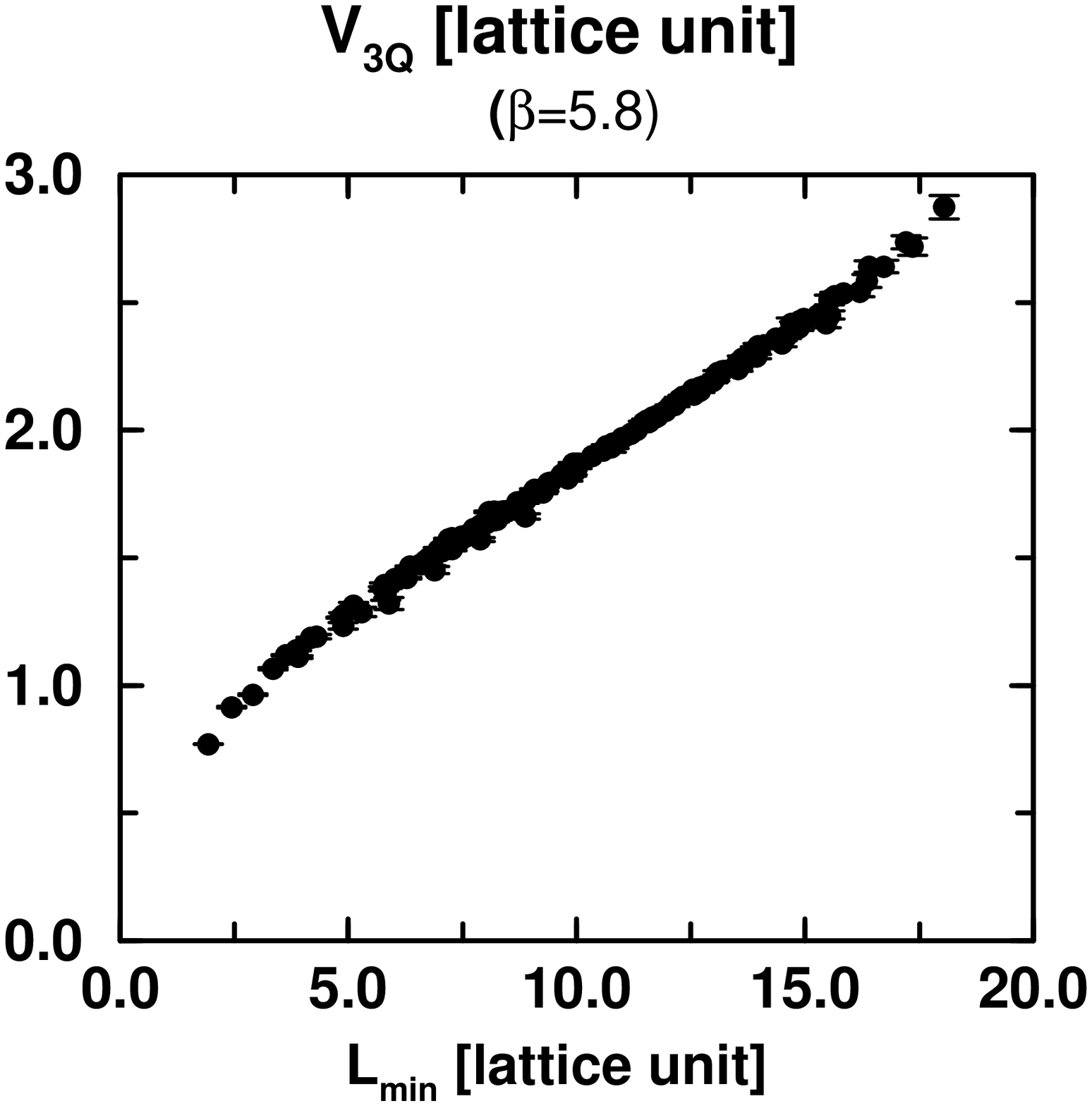,height=7cm}
\caption{
The lattice QCD data for the 3Q potential $V_{\rm 3Q}^{\rm latt}$ at $\beta=5.8$
as the function of $L_{\rm min}$, the minimum value of
the total length of the flux tubes, in the lattice unit.
}
\label{3q58nor}
\end{center}
\end{figure}

\begin{figure}[p]
\begin{center}
\epsfile{file=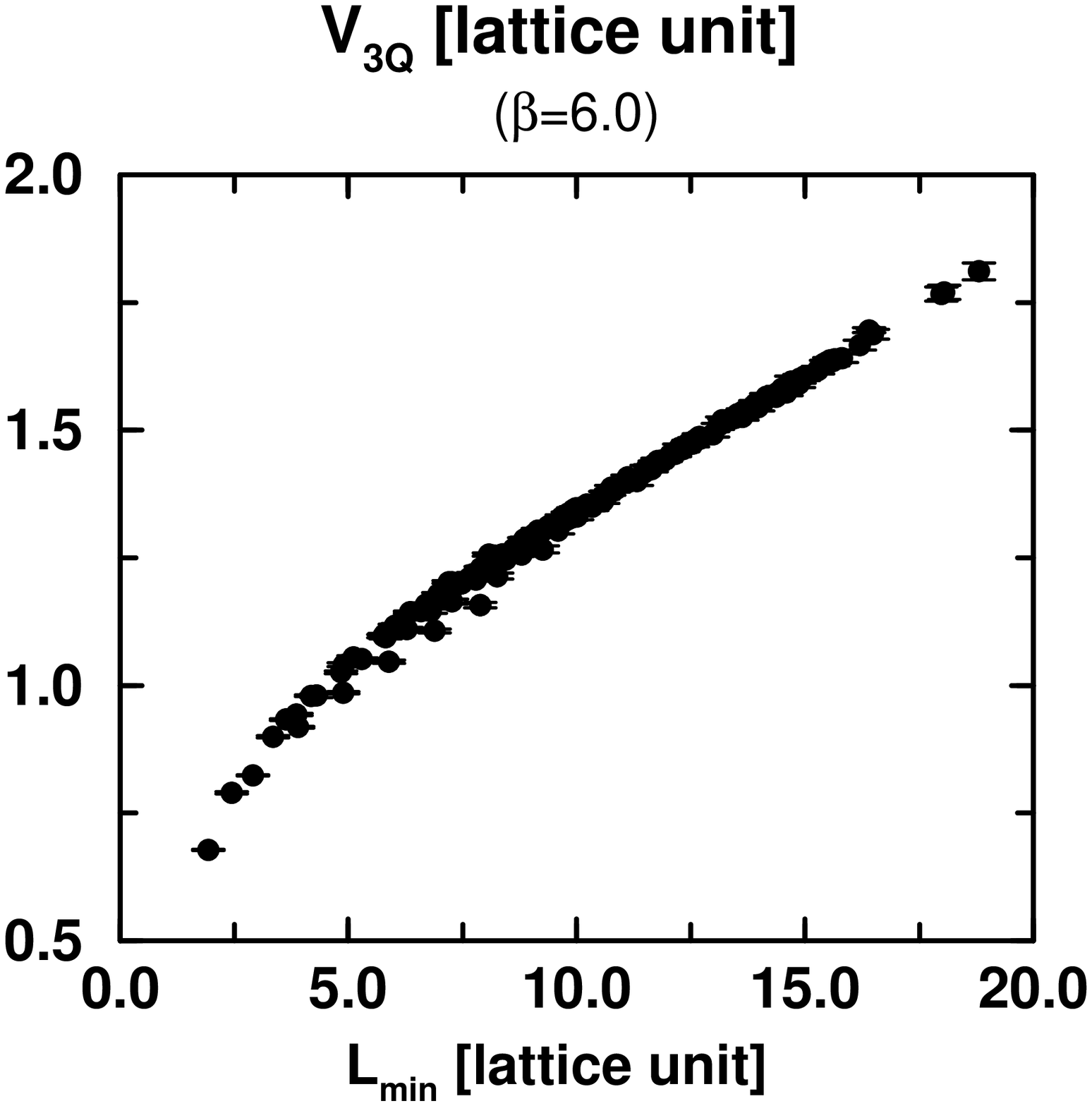,height=7cm}
\caption{
The lattice QCD data for the 3Q potential $V_{\rm 3Q}^{\rm latt}$ at $\beta=6.0$
as the function of $L_{\rm min}$, the minimum value of
the total length of the flux tubes, in the lattice unit.
}
\label{3q60nor}
\end{center}
\end{figure}

To begin with, we examine the potential form of $V_{\rm 3Q}$ at the semi-quantitative level. 
As the Q-$\bar{\rm Q}$ potential, the 3Q potential is also expected to be reproduced 
by the simple sum of the Coulomb term, the linear confinement term and a constant.
In Figs.~\ref{3q57nor}-~\ref{3q60nor}, we plot the 3Q ground-state potential $V_{\rm 3Q}$    
as the function of the minimal total flux-tube length  $L_{\rm min}$, the minimal value 
of the total length of color flux tubes linking the three quarks. as discussed in Section 2.
Apart from a constant, $V_{\rm 3Q}$ is almost proportional to $L_{\rm min}$ in the 
infrared region. 

To single out the large-distance behavior of $V_{\rm 3Q}$ by subtracting perturbative Coulomb 
contribution, we examine $V_{\rm 3Q}-V_{\rm 3Q}^{\rm Coul}$. 
Here, $V_{\rm 3Q}^{\rm Coul}$ is defined as
\begin{equation}
V_{\rm 3Q}^{\rm Coul}\equiv -\frac{A_{\rm Q\bar{Q}}}{2}
\sum_{i<j} \frac{1}{|{\bf r}_i-{\bf r}_j|}, 
\end{equation}
which is the potential form expected from the OGE process in perturbative QCD. 
Reflecting the SU(3) color factor, the coefficient in $V_{\rm 3Q}^{\rm Coul}$ between 
two quarks, of which combination belong the $\bar{3}$ representation, is set to be 
a half of the coefficient in the color-singlet Q-$\bar{\rm Q}$ system.
We note that $A_{\rm Q\bar{Q}}$ is already extracted from the lattice QCD data 
of the Q-$\bar{\rm Q}$ potential, as shown in Table~\ref{fitprm}. 

\begin{figure}[p]
\begin{center}
\epsfile{file=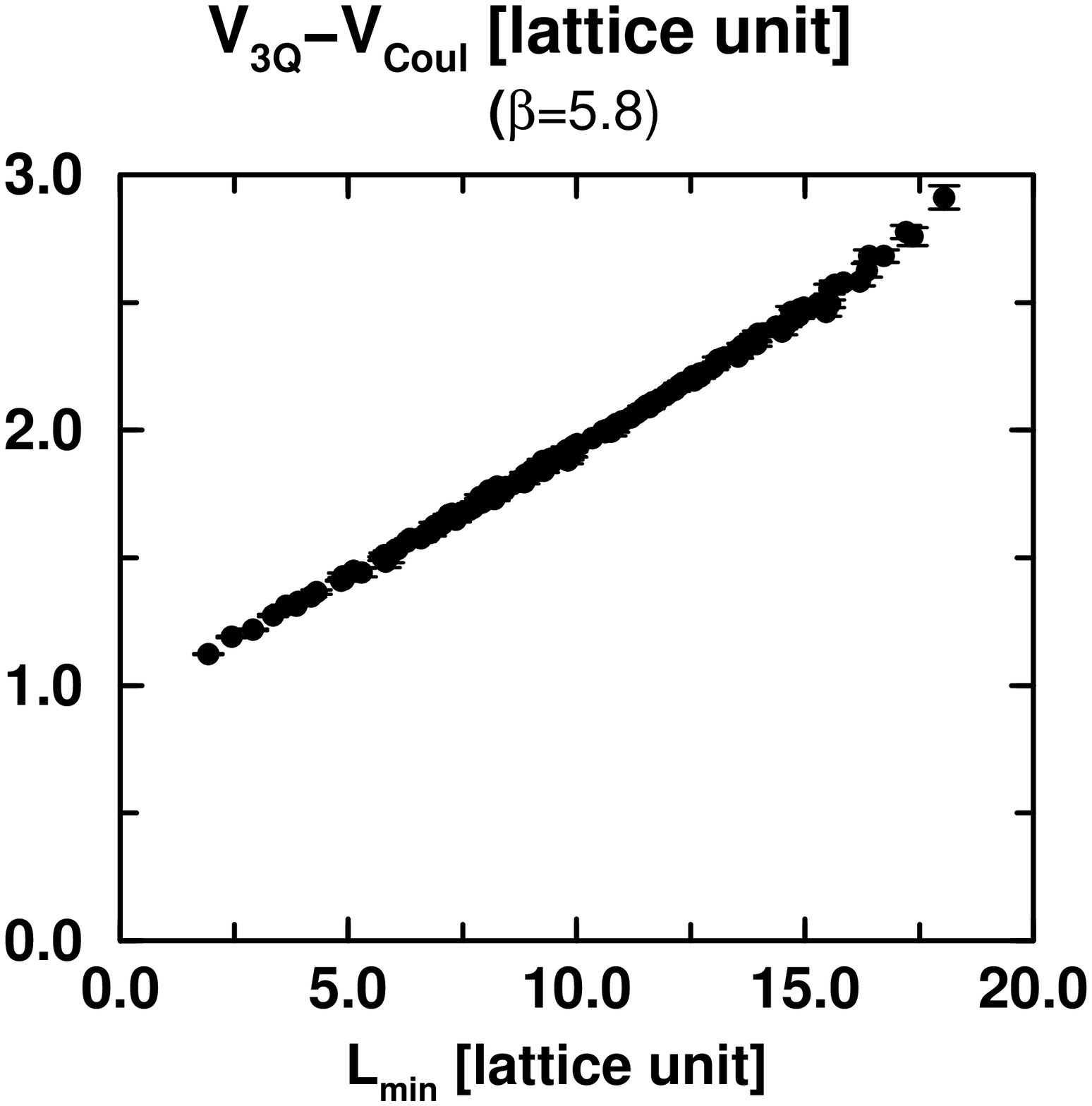,height=7cm}
\caption{
The semi-quantitative test on the confinement part in the 3Q potential
$V_{\rm 3Q}$ at $\beta=5.8$.
The Coulomb-subtracted potential $V_{\rm 3Q}^{\rm latt}-V_{\rm 3Q}^{\rm Coul}$
is plotted as the function of $L_{\rm min}$, the minimal value of the total flux-tube length.
Here, the Coulomb part $V_{\rm 3Q}^{\rm Coul}$ is evaluated from the Q-$\bar{\rm Q}$ potential.
}
\label{3q58rem}
\end{center}
\end{figure}

\begin{figure}[p]
\begin{center}
\epsfile{file=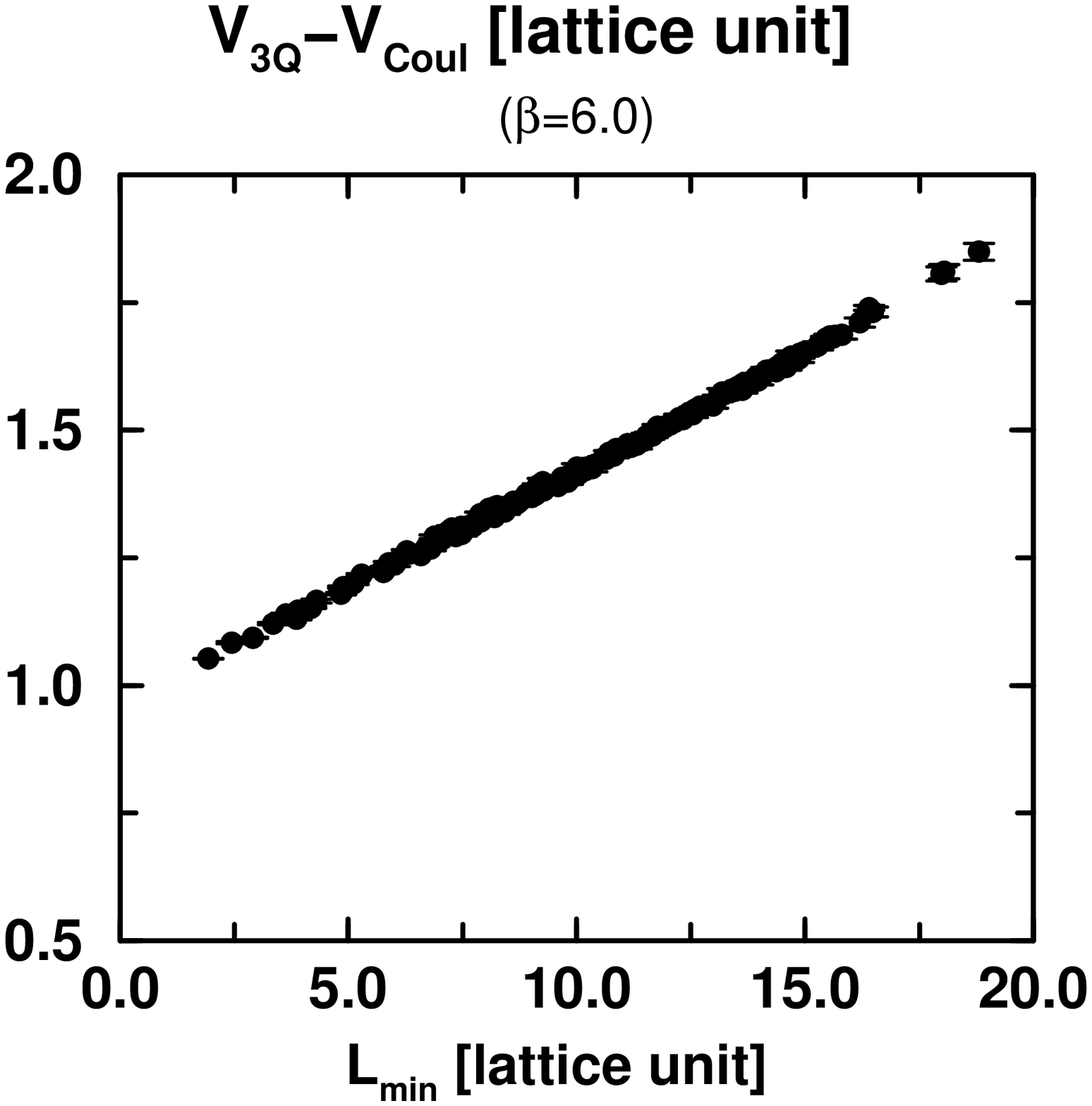,height=7cm}
\caption{
The semi-quantitative test on the confinement part in the 3Q potential
$V_{\rm 3Q}$ at $\beta=6.0$.
The Coulomb-subtracted potential $V_{\rm 3Q}^{\rm latt}-V_{\rm 3Q}^{\rm Coul}$
is plotted as the function of $L_{\rm min}$, the minimal value of the total flux-tube length.
Here, the Coulomb part $V_{\rm 3Q}^{\rm Coul}$ is evaluated from the Q-$\bar{\rm Q}$ potential.
}
\label{3q60rem}
\end{center}
\end{figure}

In Fig.~\ref{3q58rem}-\ref{3q60rem}, 
we plot $V_{\rm 3Q}-V_{\rm 3Q}^{\rm Coul}$ as a function of $L_{\rm min}$, using the lattice 
data of $V_{\rm 3Q}$ and $A_{\rm Q\bar{Q}}$ in Table~\ref{fitprm} from the Q-$\bar{\rm Q}$ 
potential. In the whole region, the linearity on $L_{\rm min}$ is observed, 
which means that the 3Q potential $V_{\rm 3Q}$ can be well described by a sum of the 
perturbative Coulomb term as $V_{\rm 3Q}^{\rm Coul}$ and the non-perturbative linear 
confinement term proportional to $L_{\rm min}$, as shown in Eq.(\ref{yansatz}). 
Thus, the lattice data seem to support the Y-ansatz. Note here that this simple fit  
is not the best fit in terms of the Y-ansatz with $(A_{\rm 3Q},\sigma_{\rm 3Q},C_{\rm 3Q})$, 
and the Y-ansatz seems to work well even in this non-best fit. In next subsection, we perform 
the fit analysis of the 3Q potential with the Y-ansatz at the quantitative level.

\subsection{The fit analysis with the Y-ansatz}
We perform the best fit analysis for the lattice QCD data of $V_{\rm 3Q}$ 
in terms of the Y-ansatz  with $(A_{\rm 3Q}, \sigma_{\rm 3Q}, C_{\rm 3Q})$ at each $\beta$.
We show in Table~\ref{fitprm} 
the best-fit parameter set $(A_{\rm 3Q}, \sigma_{\rm 3Q}, C_{\rm 3Q})$ 
in the Y-ansatz for $V_{\rm 3Q}$ at each $\beta$. 
In Table~\ref{573qpot}-\ref{603qpot4}, 
we compare the lattice data $V_{\rm 3Q}^{\rm latt}$ with the Y-ansatz 
fitting function $V_{\rm 3Q}^{\rm fit}$ in Eq.(\ref{yansatz}) 
with the best-fit parameters in Table~\ref{fitprm}.
We observe a good agreement between $V_{\rm 3Q}^{\rm latt}$ and $V_{\rm 3Q}^{\rm fit}$.
In fact, the deviation $V_{\rm 3Q}^{\rm latt}-V_{\rm 3Q}^{\rm fit}$ is only within a few \% 
of the typical scale of $V_{\rm 3Q}$ for every lattice data in 
Table~\ref{573qpot}-\ref{603qpot4}.
(Since the potential include an irrelevant constant, the typical scale of $V_{\rm 3Q}$ is to be 
understood as its typical variation among the different 3Q system rather than the value itself.)
Thus, the three-quark ground-state potential $V_{\rm 3Q}$ is well described by 
Eq.(\ref{yansatz}) of the Y-ansatz within a few \% deviation.

\begin{figure}[p]
\begin{center}
\epsfile{file=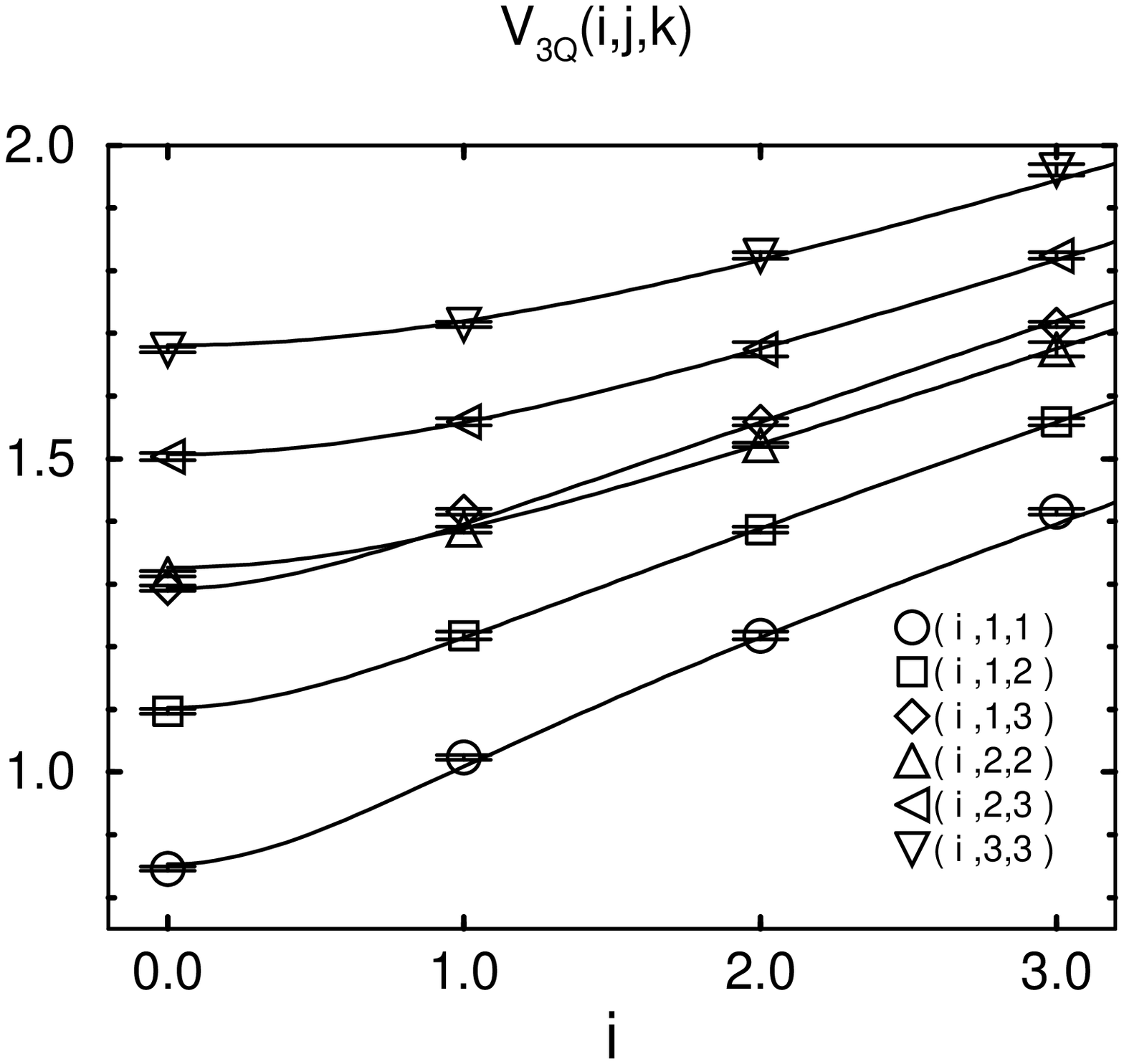,height=7cm}
\caption{
The comparison between the lattice QCD data $V_{\rm 3Q}^{\rm latt}$ at $\beta=5.7$ 
and the fitted curve of $V_{\rm 3Q}^{\rm fit}$ as the function of $i$ for each $(j, k)$ fixed, 
when the three quarks are located at $(i,0,0), (0,j,0), (0,0,k)$ in the lattice unit.
The lattice data $V_{\rm 3Q}^{\rm latt}$ are expressed as the points, 
and $V_{\rm 3Q}^{\rm fit}$ is expressed as the solid curve for each $(j, k)$.
}
\label{mplot}
\end{center}
\end{figure}

As a visual demonstration on the agreement of this fit, we compare in Fig.~\ref{mplot} 
the lattice QCD data $V_{\rm 3Q}^{\rm latt}$ at $\beta=5.7$ and the best-fit curve 
of $V_{\rm 3Q}^{\rm fit}$ as the function of $i$ for each $(j, k)$ fixed, 
when the three quarks are located at $(i,0,0), (0,j,0), (0,0,k)$ in the lattice unit.
While the lattice data $V_{\rm 3Q}^{\rm latt}$ are restricted on the integer of $i$ and are 
expressed as the points, 
$V_{\rm 3Q}^{\rm fit}$ in Eq.(\ref{yansatz}) can be calculated for arbitrary real number of $i$ 
and is expressed as a curve for each $(j, k)$.
In Fig.~\ref{mplot} at $\beta=5.7$, one finds a good agreement of the lattice QCD data 
$V_{\rm 3Q}^{\rm latt}$ and the fit curve $V_{\rm 3Q}^{\rm fit}$ for each $(j, k)$. 

In spite of the good agreement of $V_{\rm 3Q}^{\rm latt}$ with $V_{\rm 3Q}^{\rm fit}$, 
to be strict, $\chi^2/N_{\rm DF}$ listed in Table~\ref{fitprm} seems relatively large, 
which means the relatively large deviation $V_{\rm 3Q}^{\rm latt}-V_{\rm 3Q}^{\rm fit}$ 
in comparison with the error. 
Besides physical reasons, this may be due to the under-estimate of the error. 
In fact, the statistical error itself seems very small, but the error should be inevitably 
enlarged by the systematic error such as the discretization error in lattice calculations.
In particular, the statistical error for the short-distance data is rather small, and such a 
smallness of the short-distance error seems to provide the large value of $\chi^2/N_{\rm DF}$, 
which may indicate an importance to control the finite lattice-spacing effect.
Of course, this point would be clarified, if the lattice QCD study with the finer and larger 
lattice is performed. Besides the direct check on the $\beta$-dependence, the similar fit 
analysis with the lattice Coulomb potential is expected to be meaningful.
On the lattice, to be strict, the Coulomb potential is to be modified into the lattice Coulomb 
potential, which contains the finite lattice-spacing effect more directly.
Hence, the fit with the lattice Coulomb potential is expected to reduce the discretization 
error from the finite lattice spacing, especially for the short-distance data.
In the later subsection, we will perform the fit analysis using the lattice Coulomb potential. 

Finally, we compare the best-fit parameter set $(\sigma_{\rm 3Q}, A_{\rm 3Q}, C_{\rm 3Q})$ 
in the 3Q potential $V_{\rm 3Q}$ in Eq.(\ref{yansatz}) with 
$(\sigma_{\rm Q\bar{Q}}, A_{\rm Q\bar{Q}}, C_{\rm Q\bar{Q}})$ in the Q-$\bar {\rm Q}$ 
potential $V_{\rm Q\bar{Q}}$ in Eq.(\ref{qqbarfit}) as listed in Table~\ref{fitprm}. 
As a remarkable fact, we find a universal feature of the string tension, 
\begin{equation}
\sigma_{\rm 3Q} \simeq \sigma_{\rm Q\bar{Q}},
\end{equation}
as well as the OGE result for the Coulomb coefficient, 
\begin{equation}
A_{\rm 3Q} \simeq \frac12 A_{\rm Q\bar{Q}}.
\end{equation} 

\subsection{The model-independent check in the diquark limit}

As a model-independent check, we consider the diquark limit, where two quark locations 
coincide in the 3Q system. In the diquark limit, the static 3Q system becomes equivalent to 
the Q-$\bar{\rm Q}$ system, which leads to a physical requirement on the relation between 
$V_{\rm 3Q}$ and $V_{\rm Q \bar{Q}}$. Our results, $\sigma_{\rm 3Q} \simeq 
\sigma_{\rm Q\bar{Q}}$ and $A_{\rm 3Q} \simeq \frac12 A_{\rm Q\bar{Q}}$, 
are consistent with the physical requirement in the diquark limit. 

Next, we consider the constant terms $C_{\rm 3Q}$ in the diquark limit, although 
such a constant term is a lattice artifact and is physically irrelevant.
As a caution in the continuum diquark limit, there appears a singularity or a divergence 
from the Coulomb term in $V_{\rm 3Q}$ as 
\begin{equation}
\lim_{{\bf r}_j \rightarrow {\bf r}_i} \frac{-A_{\rm 3Q}}{|{\bf r}_i-{\bf r}_j|}=-\infty. 
\end{equation}
In the lattice regularization, this ultraviolet divergence is regularized to be a 
finite constant with the lattice spacing $a$ as 
\begin{equation}
\frac{-A_{\rm 3Q}}{|{\bf r}_i-{\bf r}_j|} \rightarrow \frac{-A_{\rm 3Q}}{\omega a},
\end{equation}
where $\omega$ is a dimensionless constant satisfying $0 < \omega <1$ and $\omega \sim 1$. 
Then, we find 
\begin{equation}
C_{\rm 3Q}+\frac{-A_{\rm 3Q}}{\omega a}=C_{\rm Q\bar{Q}}, 
\end{equation}
or equivalently, 
\begin{equation}
C_{\rm 3Q}-C_{\rm Q\bar{Q}}=\frac{A_{\rm 3Q}}{\omega a} \quad ( > 0)
\end{equation}
in the diquark limit.  
This is the requirement for the constant term in the diquark limit on the lattice. 
Our lattice QCD results for $C_{\rm 3Q}$, $C_{\rm Q \bar{Q}}$ and $A_{\rm 3Q}$ are thus 
consistent with this requirement, and one finds $\omega \simeq 0.41-0.45$:  
$\omega(\beta=5.7) \simeq 0.447$, $\omega(\beta=5.8) \simeq 0.414$, 
$\omega(\beta=6.0) \simeq 0.424$. 

\subsection{The Y-ansatz with the Yukawa potential}

In the previous subsection, we adopt the Coulomb potential as the short-distance ingredient, 
because the OGE process is expected to be dominant at the short distance and 
the Q-$\bar{\rm Q}$ potential seems to be reproduced with the Coulomb plus linear 
potential in the lattice QCD. 
The first reason is, however, nontrivial in the intermediate and the infrared regions, 
where the perturbative QCD would not work.
In fact, due to some nonperturbative effects besides the confinement potential, 
the Coulomb potential caused by the OGE process may be modified in the infrared region. 

For instance, the dual superconductor theory for the 
quark confinement\cite{SST95,SSTI95,CH95,QC01} supports the Yukawa plus linear potential 
rather than the Coulomb plus linear potential, although the dual gluon mass $m_B$ 
appearing in the exponent in the Yukawa potential 
may not be so large, e.g., $m_B \sim 0.5 {\rm GeV}$ both in the model framework
\cite{SST95,SSTI95} and in the lattice study\cite{SAIT00,SITA98}.

From the theoretical viewpoint, such a possibility on the infrared screening of the 
Coulomb potential seems rather attractive in terms  of the empirical absence of 
the color Van-Der-Waals force in the infrared limit\cite{FS79}.
In fact, if the two-body Coulomb potential is not screened in the infrared limit, 
the color Van-Der-Waals force inevitably appears as a long-distance force between hadrons, 
which is not observed experimentally.

Then, we also investigate the fit analysis of $V_{\rm 3Q}$ with the 
Y-ansatz with the Yukawa potential as 
\begin{equation}
V_{\rm 3Q}^{\rm Yukawa} \equiv 
-A_{\rm 3Q}^{\rm Yukawa}\sum_{i<j}V^{\rm Yukawa}(|{\bf r}_i-{\bf r}_j|)+
\sigma^{\rm Yukawa}_{\rm 3Q} L_{\rm min}+C^{\rm Yukawa}_{\rm 3Q}, 
\end{equation}
where $V^{\rm Yukawa}(r)$ denotes the normalized Yukawa potential 
\begin{equation}
V^{\rm Yukawa}(r) \equiv \frac1r e^{-m_Br}.
\end{equation}
Here, $m_B$ corresponds to the dual gluon mass in the dual superconductor picture 
\cite{N74,SST95,SSTI95}. 

We find that the best-fit analysis of the lattice QCD data $V_{\rm 3Q}^{\rm latt}$ 
with $V_{\rm 3Q}^{\rm Yukawa}$ indicates $m_B \simeq 0$.
Of course, in this special case of $m_B \simeq 0$, the Yukawa potential reduces 
the Coulomb potential, and the result almost coincides with that in the previous fit.
Thus, through the fit analysis with the Yukawa potential based on the Y-ansatz, 
we have observed no definite evidence to replace the Coulomb-potential part 
by the Yukawa potential in the present calculation.

\subsection{The Y-ansatz with the lattice Coulomb potential}
So far, we have investigated the fit analysis of the 3Q potential $V_{\rm 3Q}$ mainly with 
the continuum Coulomb and the confinement potential. 
However, in comparing with the lattice data, the careful treatment considering the lattice 
discretization effect may be desired. The main effect of the lattice discretization 
appears only at the short distance, and hence no modification would be necessary for the 
confinement potential, which becomes significant only at the large distance. 
On the other hand, it is nontrivial to use the continuum Coulomb potential for the fit of 
the lattice data, especially at the short distance, according to the lattice discretization.
For instance, the short-distance singularity of the Coulomb potential becomes rather 
smeared on the lattice, as will be shown later.

In this subsection, we perform the fit analysis of the lattice QCD data of $V_{\rm 3Q}$ 
in terms of the Y-ansatz with the lattice Coulomb potential.
The lattice Coulomb (LC) potential between two color charges is obtained with the lattice 
Coulomb propagator as
\begin{equation}
V^{\rm LC}(\vec{n}) \equiv \pi\int_{-\pi/a}^{\pi/a}\frac{d^3q}{(2\pi)^3}
\frac{\exp(-i\vec{p}\cdot\vec{n}a)}{\sum_{i=1}^{3}\sin ^2(p_ia/2)},
\label{lattcoul}
\end{equation}
where $\vec{n}\equiv (n_1,n_2,n_3) \in {\bf Z}^3$ denotes the relative vector between 
the two color charges in the lattice unit.
Here, the lattice Coulomb potential $V^{\rm LC}(\vec{n})$ is properly normalized 
so as to reduce into the 1/r potential in the continuum limit as 
\begin{equation}
V^{\rm LC}(\vec{n}) \rightarrow \frac{1}{r}
\end{equation}
with $r = |\vec n| a$.

\begin{figure}[p]
\begin{center}
\epsfile{file=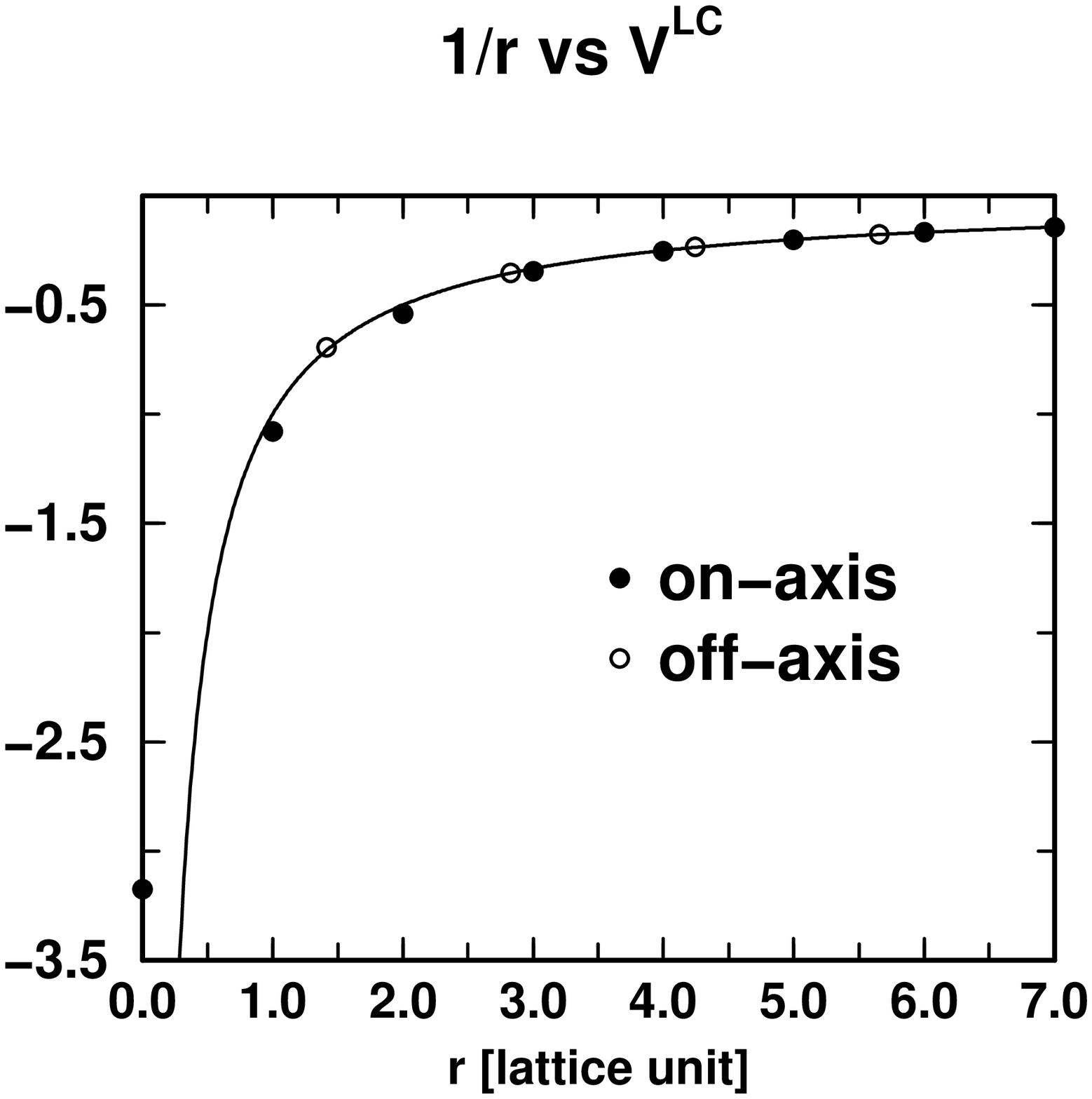,height=7cm}
\caption{
The comparison of the lattice Coulomb (LC) potential $V^{\rm LC}(\vec n)$
with $1/r$ in the lattice unit.
We plot the on-axis data $V^{\rm LC}(k,0,0)$ ($0\leq k\leq7$) by the closed circle,
the off-axis data $V^{\rm LC}(k,k,0)$ ($1\leq k\leq4$) by the open circle,
and $1/r$ by the solid curve.
}
\label{compar}
\end{center}
\end{figure}

In Fig.~\ref{compar}, we plot $V^{\rm LC}(\vec{n})$ as a function of $|\vec{n}|$
together with $1/r$ as a function of $r$.
One find that the singularity near the origin $|\vec{n}|=0$ becomes smeared, and 
$V_{\rm LC}$ takes a finite value even at $\vec{n}=\vec 0$, 
which was mentioned in the context of the diquark limit in the previous subsection.
This may cause a significant deviation in the fit analysis with the $1/r$ Coulomb 
potential, especially for the short-distance data analysis.

To begin with, we examine the fit analysis of the lattice QCD data on the Q-$\bar {\rm Q}$ 
potential $V_{\rm Q\bar{Q}}$ with 
\begin{equation}
V_{\rm Q\bar{Q}}^{LC} \equiv -A_{\rm Q\bar{Q}}^{\rm LC} V^{\rm LC}(\vec{n})
+\sigma_{\rm Q\bar{Q}}^{\rm LC}|\vec{n}|+C_{\rm Q\bar{Q}}^{\rm LC}, 
\end{equation}
using the lattice Coulomb potential $V^{\rm LC}(\vec{n})$. We refer to this fit as the LC fit. 
We list in Table~\ref{fitprm} the best-fit parameter set ($A_{\rm Q\bar{Q}}^{\rm LC}$, 
$\sigma^{\rm LC}_{\rm Q\bar{Q}}$, $C^{\rm LC}_{\rm Q\bar{Q}}$) at each $\beta$. 
In Table~\ref{fitprm}, the label ``on-axis" means the fit analysis for the on-axis data only, 
and the label ``off-axis" means the fit analysis for both on-axis and off-axis data. 
In the LC fit, a significant reduction of $\chi^2/N_{\rm DF}$ is observed, 
in comparison with the fit with the $1/r$ Coulomb potential in Eq.(\ref{qqbarfit}),
as shown in Table~\ref{fitprm}, in spite of the similar values of the fit parameters.
We find a further good agreement between the lattice QCD data and 
the fitted values of $V_{\rm Q \bar Q}^{\rm LC}$ on the Q-$\bar{\rm Q}$ potential. 
In particular, the LC fit shows an acceptable value of $\chi^2/N_{\rm DF}$ 
even for the fit on both on-axis and off-axis data, while the fit with the $1/r$ 
Coulomb potential shows the extremely large $\chi^2/N_{\rm DF}$ about 10 
when the off-axis data are included. 
Thus, the discretization effect, which may not be negligible for short-distance data, 
seems to be taken into account neatly by the use of the LC potential to some extent. 
Accordingly, the fit analysis with the lattice Coulomb potential $V^{\rm LC}$ seems to 
provide a more precise information also for the linear confinement potential,
although the string tension obtained from the fit analysis is almost unchanged as  
$\sigma_{\rm Q \bar Q}^{\rm LC} \simeq \sigma_{\rm Q \bar Q}$.

Now, we perform the LC fit analysis of the lattice data on the 3Q potential $V_{\rm 3Q}$ with 
\begin{equation}
V_{\rm 3Q}^{\rm LC} \equiv -A_{\rm 3Q}^{\rm LC}\sum_{i<j}V^{\rm LC}(\vec{n_{ij}})+
\sigma^{\rm LC}_{\rm 3Q} L_{\rm min}+C^{\rm LC}_{\rm 3Q}, 
\label{LCyansatz}
\end{equation}
using the lattice Coulomb potential $V^{\rm LC}(\vec{n})$. 
In Table~\ref{fitprm}, we list the best-fit parameter set ($A_{\rm 3Q}^{\rm LC}$, 
$\sigma^{\rm LC}_{\rm 3Q}$,$C^{\rm LC}_{\rm 3Q}$) together with the best-fit parameters
($A_{\rm Q\bar{Q}}^{\rm LC}$, $\sigma^{\rm LC}_{\rm Q\bar{Q}}$, $C^{\rm LC}_{\rm Q\bar{Q}}$).
In this LC fit, $\chi^2/N_{\rm DF}$ is reduced in comparison with the fit with 
the $1/r$ Coulomb potential at each $\beta$, as shown in Table~\ref{fitprm}. 
Then, this fit seems to be acceptably good even without taking account of 
the remaining systematic error.  
Accordingly, the 3Q potential data are well reproduced with the Y-ansatz fit function 
$V_{\rm 3Q}^{\rm LC}$ in Eq.(\ref{LCyansatz}) 
with the lattice Coulomb potential with accuracy better than 1 \% .
Again, the fit analysis with the lattice Coulomb potential $V^{\rm LC}$ is expected to 
control the discretization effect to some extent, and would provide a more precise 
information also for the linear confinement potential. 

Finally, we focus on the best-fit parameter set $(\sigma_{\rm 3Q}^{\rm LC}, A_{\rm 3Q}^{\rm LC}, 
C_{\rm 3Q}^{\rm LC})$ in the Y-ansatz with the lattice Coulomb potential. 
The values of the best-fit parameters in the LC fit are almost the same as 
those in the previous fit with the $1/r$ Coulomb potential as 
\begin{equation}
(\sigma_{\rm 3Q}^{\rm LC}, A_{\rm 3Q}^{\rm LC}, C_{\rm 3Q}^{\rm LC}) 
\simeq (\sigma_{\rm 3Q}, A_{\rm 3Q}, C_{\rm 3Q}).
\end{equation} 
In particular, the string tension obtained from the fit analysis is almost unchanged as  
$\sigma_{\rm 3Q}^{\rm LC} \simeq \sigma_{\rm 3Q}$.
We then find again the universality of the string tension as $\sigma^{\rm LC}_{\rm 3Q} \simeq 
\sigma^{\rm LC}_{\rm Q\bar{Q}}$ at each $\beta$. 
The OGE relation on the Coulomb coefficient is found as 
$A_{\rm 3Q}^{\rm LC} \simeq \frac12 {A_{\rm Q\bar{Q}}^{\rm LC}}$.
In particular, this OGE relation seems to be observed precisely at $\beta$=6.0, which 
is the finest and the most reliable lattice in the present calculation. 
(At $\beta$=5.7, 5.8, the ratio ${A_{\rm 3Q}^{\rm LC}}/{A_{\rm Q\bar{Q}}^{\rm LC}}$
is about 0.4, which seems slightly different from $\frac12$ as the OGE result.
Note that, however, the discretization error on the QCD action still remains at the small 
$\beta$ even with the lattice Coulomb potential. In addition, at $\beta=5.7, 5.8$, 
the nearest site is relatively far, and hence the Coulomb contribution to the 3Q potential 
is relatively small, which may lead to an uncertainty of the Coulomb coefficient.) 

To summarize this section, we conclude that the lattice QCD data of the 3Q potential 
$V_{\rm 3Q}$ can be fairly described with the Y-ansatz within a few \% deviation, 
and therefore the nonperturbative linear confinement potential 
is proportional to $L_{\rm min}$, the minimal value of the total length of the Y-type 
flux tube linking the three quarks, which supports the Y-ansatz. 

\section{Comparison with the delta ansatz and the generalized Y-ansatz}
\subsection{Comparison with the Delta ansatz}
For the 3Q potential, the $\Delta$-ansatz is also an interesting candidate~\cite{C96}
as well as the Y-ansatz. The $\Delta$-ansatz is expressed as
\begin{equation}
V_{\rm 3Q}=-A_\Delta \sum_{i<j}\frac1{|{\bf r}_i-{\bf r}_j|}
+\sigma_\Delta \sum_{i<j} |{\bf r}_i-{\bf r}_j|+C_\Delta,
\label{dansatz}
\end{equation}
which consists of the two-body linear potential between quarks. This $\Delta$-ansatz has been 
adopted in a simple nonrelativistic quark potential model~\cite{OY81}, 
because of its simplicity for the calculation.
In addition, several lattice QCD studies for the 3Q potential 
supported the $\Delta$ ansatz~\cite{SW8486,TES88,B01,AFT01,AFT02}.
However, Refs.~\cite{SW8486,TES88} seem rather old done in 14 years ago, 
and were done without smearing, so that the excited-state potential 
may largely contribute in their measurements. (See Fig.~\ref{czero}.)
In recent paper as Ref.~\cite{AFT02}, in spite of the use of the smearing, the authors simply 
compared the 3Q potential $V_{\rm 3Q}$ with the Y and the $\Delta$ ans\"atze 
with a fixed string tension estimated from the Q-$\bar{\rm Q}$ potential, 
only for several equilateral-triangle 3Q configurations without the quantitative fit analysis.
Furthermore, in Refs.~\cite{B01,AFT02}, the ``deviation'' between the lattice data and 
the Y-ansatz seems to be explained as a trivial constant shift of the potential.
In fact, the potential calculated in lattice QCD includes a physically irrelevant constant, 
which is not properly scaled in the physical unit, as shown in Table~\ref{fitprm}.
If such an irrelevant constant is properly removed,  
the Y-ansatz seems to be better than $\Delta$-ansatz even in Refs.~\cite{B01,AFT01,AFT02}, 
e.g., the slope of the lattice data $V_{\rm 3Q}^{\rm latt}$ seems closer to the 
Y-ansatz rather than the $\Delta$-ansatz.
In Ref.~\cite{AFT02}, the authors set the potential origin at the nearest lattice point 
and used the continuum Coulomb potential for the fit function, but this seems dangerous 
because a constant deviation may appear as a mismatch between the lattice Coulomb 
potential which is preferable to the lattice data and the continuum Coulomb potential 
for the fit function, as shown in Fig.~\ref{compar}. 
In any case, the quantitative fit analysis is essential for the study of the 
functional form of the 3Q potential.

In this subsection, we perform the fit analysis with the $\Delta$ ansatz.
To begin with, we try to fit $V_{\rm 3Q}^{\rm latt}$ with the $\Delta$-ansatz in Eq.(\ref{dansatz}),
which was suggested in Refs.\cite{SW8486,TES88,C96}.
We list in Table~\ref{dfitprm} the best fit parameter set  
$(A_\Delta, \sigma_\Delta, C_\Delta)$ in the $\Delta$-ansatz at each $\beta$. 
In comparison with the Y-ansatz, this fit with the $\Delta$-ansatz seems rather 
worse, because of the larger value of $\chi^2/N_{\rm DF}$. 
In fact, $\chi^2/N_{\rm DF}$ is unacceptably large as $\chi^2/N_{\rm DF}=10.1$ at $\beta=5.7$, 
$\chi^2/N_{\rm DF}=13.7$ at $\beta=5.8$ even for the best fit. 
(Of course, when $\sigma_{\rm \Delta}$ is fixed to be the half value 
of the string tension
in 
the Q-$\bar{\rm Q}$ potential as in Refs.\cite{TES88,B01,AFT01,AFT02}, 
this fit with the $\Delta$-ansatz becomes further worse with a larger $\chi^2/N_{\rm DF}$.)

As an approximation, however, $V_{\rm 3Q}$ seems described by a simple sum of 
the effective two-body Q-Q potential with a reduced string tension as 
\begin{equation}
\sigma_\Delta \simeq 0.53  \sigma_{\rm Q \bar Q}.
\label{deltastring}
\end{equation} 
This reduction factor can be naturally understood as a geometrical factor rather than the 
color factor, since the ratio between $L_{\rm min}$ and the perimeter length $L_P$ 
of the 3Q triangle satisfies 
\begin{equation}
\frac12 \le \frac{L_{\rm min}}{L_P} \le \frac1{\sqrt{3}},
\end{equation}
which leads to $L_{\rm min} \sigma = L_P \sigma_{\rm QQ}$ 
with $\sigma_{\rm QQ}=(0.5 \sim 0.58) \sigma$. 
The OGE relation is also found as $A_\Delta \simeq \frac12 A_{\rm Q \bar Q}$.

For the fair comparison, we also examine the fit analysis by $\Delta$-ansatz
with the lattice Coulomb potential, as was done for the Y-ansatz in Section 5-5.
The 3Q data are fitted by the form of 
\begin{equation}
V_{\rm 3Q}=-A_\Delta^{\rm LC} \sum_{i<j}V^{\rm LC}(\vec{n_{ij}})
+\sigma_\Delta^{\rm LC} \sum_{i<j} |{\bf r}_i-{\bf r}_j|+C_\Delta^{\rm LC}.
\end{equation}
We add the results in Table~\ref{dfitprm}.
Again, a reduction in $\chi^2/N_{\rm DF}$ is observed. 
The values of the best-fit parameters are almost unchanged, and hence 
Eq.(\ref{deltastring}) and the OGE relation also hold.
However, in comparison with the Y-ansatz with the lattice Coulomb potential, 
$\chi^2/N_{\rm DF}$ in this fit with the $\Delta$-ansatz is still larger,   
and therefore the $\Delta$-ansatz is difficult to be accepted.

\subsection{A more general ansatz - the generalized Y-ansatz}

 From the theoretical reason of the short-distance perturbative QCD and the large-distance 
strong-coupling QCD, the Y-ansatz seems reasonable in the both limits. 
The overall lattice QCD data for the 3Q potential also support the Y-ansatz rather 
than $\Delta$ ansatz. However, it is not trivial whether the Y-ansatz holds 
in the intermediate region as $0.2 \ {\rm fm} < r_{ij} < 0.8 \ {\rm fm}$. 
In fact, as was conjectured by Conwall\cite{C96}, there is a possibility of 
the $\Delta$-ansatz contamination in this region, where strong-coupling QCD 
is not applicable. In addition, a few recent lattice works seem to 
support the $\Delta$-ansatz for the 3Q potential in the intermediate region, 
although their studies were performed only for the equilateral triangle 3Q configuration. 
Of course, it is rather difficult to analyze the short distance behavior of 
the non-Coulomb part of the 3Q potential, because the Coulomb part is dominant there. 
Furthermore, the Coulomb potential form itself is no more trivial 
in the intermediate region as $0.2 \ {\rm fm} < r_{ij} <0.8 \ {\rm fm}$, 
where the perturbative QCD would not be valid.

In this section, we investigate the lattice QCD data for the 3Q potential using a more 
general ansatz which includes both the Y and the $\Delta$ ans\"atze in some limits.  
On the adoption of the general ansatz, we consider the possibility of the flux-tube core 
effect. For instance, in the dual superconductor picture\cite{SST95,SSTI95,CH95,QC01}, 
the hadron flux tube has an intrinsic structure of the core region inside, 
In fact, if there exists the flux-tube core with the cylindrical radius $R_{\rm core}$, 
the Y-type flux tube may be almost identical to the $\Delta$-type configuration 
at the short distance as $r_{ij} \sim R_{\rm core}$, and thus the flux-tube length 
becomes obscured there.

\begin{figure}[h]
\begin{center}
\epsfile{file=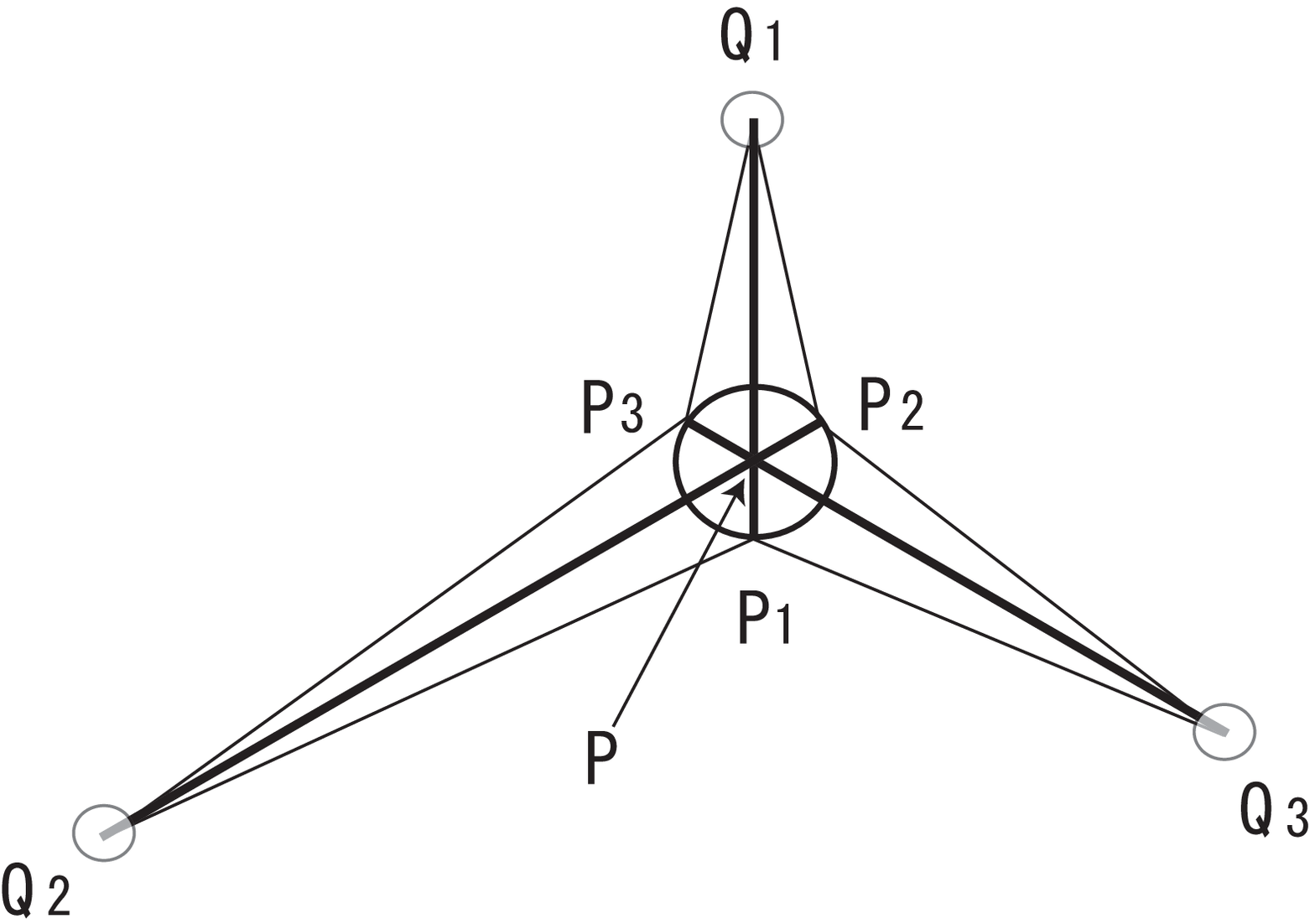,height=6cm}
\caption{
The visual illustration of the generalized Y-ansatz.
The three quarks are spatially fixed at $Q_1$, $Q_2$ and $Q_3$.
The circle around the Fermat point P corresponds to the flux-tube core
around the physical junction.
The point ${\rm P}_1$ is taken inside the circle so as to minimize 
${\rm P}_1{\rm Q}_2+{\rm P}_1{\rm Q}_3$, and ${\rm P}_2$ and ${\rm P}_3$ 
are similarly defined.
}
\label{gY}
\end{center}
\end{figure}

So far, we have defined the minimal flux-tube length, which is proportional to the 
linear potential, as 
\begin{equation}
L_{\rm min}\equiv PQ_1+PQ_2+PQ_3 
\end{equation}
with the Fermat point $P$ in Fig.~\ref{gY}. In this section, considering a possible 
flux-tube core effect, we introduce the modified minimal flux-tube length 
$\bar L_{\rm min}$ defined as 
\begin{equation}
\overline{L_{\rm min}} \equiv \frac12 
(\overline{Q_1P_2Q_3} + \overline{Q_2P_3Q_1} + \overline{Q_3P_1Q_2}),
\end{equation} 
with $\overline{ABC} \equiv \overline{AB} + \overline{BC}$, 
as shown in Fig.~\ref{gY}.
Here, the points $P_k (k=1,2,3)$ are taken inside the circle centered at the Fermat 
point $P$ with the radius $R_{\rm core}$, and $P_1$ is chosen so as to minimize 
\begin{equation}
\overline{Q_3P_1Q_2}\equiv \overline{P_1Q_2}+\overline{P_1Q_3}
\end{equation}
and so on. 
In this definition, when the circle crosses or includes the line $Q_2Q_3$, 
$P_1$ can be taken on the line $Q_2Q_3$ and then one finds 
\begin{equation}
\overline{Q_3P_1Q_2} =\overline{Q_2Q_3}.
\end{equation}
It is worth mentioning that there are two special cases corresponding 
to the Y and the $\Delta$ ans\"atze: 
\begin{equation}
\overline{L_{\rm min}}=L_{\rm min}
\end{equation}
in the case of $R_{\rm core}=0$ or $R_{\rm core} \ll r_{ij}$, and 
\begin{equation}
\overline{L_{\rm min}} =\frac12 
(\overline{Q_1Q_2} +\overline{Q_2Q_3} +\overline{Q_3Q_1}) 
\end{equation}
in the limit of $R_{\rm core}=\infty$ or $R_{\rm core} \gg r_{ij}$. 

Using this modified minimal flux-tube length $\overline L_{\rm min}$, we define the 
generalized Y-ansatz as 
\begin{equation}
V_{\rm 3Q}=\sigma_{\rm GY} \overline{L_{\rm min}}-
A_{\rm GY} \sum_{i<j} \frac1{|{\bf r}_i-{\bf r}_j|} +C_{\rm GY}.
\end {equation}
This generalized Y-ansatz includes both the Y and the $\Delta$ ans\"atze in the special 
cases of $R_{\rm core}=0$ and $R_{\rm core}=\infty$, respectively. In fact, the generalized 
Y-ansatz reproduces the Y-ansatz at the large distance, which is theoretically supported by 
strong-coupling QCD, with including the $\Delta$-ansatz behavior at the short distance. 

Then, we investigate the fit analysis for the lattice QCD data of the 3Q potential 
using the generalized Y-ansatz with varying $R_{\rm core}$. 
We show in Table~\ref{gyfitprm} the result of the fit analysis at $\beta$=5.8 and 6.0. 
We observe the best fitting at $R_{\rm core} \simeq 0.08\ {\rm fm}$ both at $\beta$=5.8 and 6.0. 
The values of the best-fit parameters $(\sigma_{\rm GY}$, $A_{\rm GY}$, $C_{\rm GY})$ 
are almost the same as those in the Y-ansatz. 
This result also seems to support the Y-ansatz rather than the $\Delta$ ansatz at the 
hadronic scale as $r \gg 0.1{\rm fm}$. 
(As an interesting speculation, $R_{\rm core} \simeq 0.08\ {\rm fm}$, which is 
almost the same both at $\beta=5.8$ and 6.0,  
may physically correspond to the flux-tube core radius in the dual superconductor picture.)

\section{Summary and Concluding Remarks}
We have studied the static three-quark (3Q) potential in detail using SU(3) lattice QCD
with $12^3 \times 24$ at $\beta=5.7$ and $16^3 \times 32$ at $\beta=5.8, 6.0$ at the
quenched level. In the first half of this paper, we have performed accurate measurement of
the 3Q Wilson loop with the smearing technique, which reduces excited-state contaminations,
and have presented the lattice QCD data of the 3Q ground-state potential $V_{\rm 3Q}$
for more than 300 different patterns of the 3Q systems.

In the latter half, we have investigated the fit analysis on $V_{\rm 3Q}$,
and have found that the lattice QCD data of the 3Q potential $V_{\rm 3Q}$ are well reproduced
within a few \% deviation by the sum of a constant, the two-body Coulomb term
and the three-body linear confinement term $\sigma_{\rm 3Q} L_{\rm min}$,
with $L_{\rm min}$ the minimal value of the total length of color flux tubes linking the
three quarks. We have investigated also the fit analysis with the lattice Coulomb potential
instead of the Coulomb potential, and have found a better fit with keeping the similar result.
 From the comparison with the Q-$\bar {\rm Q}$ potential, we have found a universality of
the string tension as $\sigma_{\rm 3Q} \simeq \sigma_{\rm Q \bar Q}$ and the
one-gluon-exchange result for the Coulomb coefficients as
$A_{\rm 3Q} \simeq \frac12 A_{\rm Q \bar Q}$.

We have also performed the various fit analyses.
Through the fit with the Yukawa potential based on the Y-ansatz, we have observed
no definite evidence that the short-distance potential becomes the Yukawa potential.
The fit with the $\Delta$-ansatz is worse than that with the Y-ansatz
on the confinement part in the 3Q potential $V_{\rm 3Q}$, although $V_{\rm 3Q}$
seems to be approximated by the $\Delta$-ansatz with $\sigma_\Delta \simeq 0.53 \sigma$.
We have considered a more general ansatz including the Y and the $\Delta$ ans\"atze
in some limits, and have found a possibility that the Y-type flux tube
has a flux-tube core about 0.08 fm, which may appear as a small mixing of the
$\Delta$-ansatz at the short distance,
although such a small $\Delta$-type contamination is negligible
in the intermediate and the infrared regions.
To conclude, all of these detailed fit analyses for the lattice QCD data of
the 3Q potential support the Y-ansatz.

\section*{Acknowledgement}
We thank Dr. T.~Umeda for his useful comments on the programming technique. 
H.S. is supported by Grant for Scientific Research (No.12640274)
from Ministry of Education, Culture, Science and Technology, Japan.
H.~M. is supported by Japan Society for the Promotion of Science
for Young Scientists. The lattice QCD Monte Carlo calculations 
have been performed on NEC-SX4 and NEC-SX5 at Osaka University and on HITACHI-SR8000 at KEK.

%\end{document}

\begin{figure}[p]
\begin{center}
\epsfile{file=lmin.eps,height=4cm}
\caption{The flux-tube configuration of the 3Q system with the minimal value of the total
flux-tube length. There appears a physical junction linking the three flux tubes at
the Fermat point P.}
\label{lmin}
\end{center}
\end{figure}

\newpage
%%%%%%%%%%%%%%%%%%%%%%%%%%%%%%%%%%%%%%%%%%%%%%%%%%%%%%%%%%%%%%%%%
\begin{table}[h]
\vspace{0.2cm}
\newcommand{\m}{\hphantom{$-$}}
\newcommand{\cc}[1]{\multicolumn{1}{c}{#1}}
\begin{center}
\begin{tabular}{ c c c c c c c c c c } \hline\hline
$\beta$ & $a$[fm] & lattice size & 
$N_{\rm 3Q}$ & $N_{\rm conf}$ &
$N_{\rm therm}$ & $N_{\rm sep}$ &  $\alpha$ & $N_{\rm smr}$ &super-computer
\\ \hline
5.7 & 0.19 & $12^3\times 24$ &  16 & 210 &  5,000 & 500 & 2.3 & 12 & NEC SX-4 \\
5.8 & 0.14 & $16^3\times 32$ & 139 & 200 & 10,000 & 500 & 2.3 & 22 & HITACHI SR8000 \\
6.0 & 0.10 & $16^3\times 32$ & 155 & 150 & 10,000 & 500 & 2.3 & 42 & NEC SX-5
\\ 
\hline\hline
\end{tabular}
\end{center}
\caption{
The simulation condition and the related information.
For each $\beta$, the corresponding lattice spacing $a$, the lattice size, 
the number $N_{\rm 3Q}$ of the different patterns of the 3Q system analyzed, 
the number $N_{\rm conf}$ of the gauge configuration used for the measurement, 
the number $N_{\rm therm}$ of sweeps for the thermalization,
the number $N_{\rm sep}$ of sweeps for the separation,
the smearing parameter $\alpha$, 
the iteration number $N_{\rm smr}$ of the suitable smearing for the 3Q potential, 
and the used super-computer are listed.}
\label{gaugeprm}
\end{table}

\normalsize
\begin{table}[h]
\begin{center}
\newcommand{\m}{\hphantom{$-$}}
\newcommand{\s}{\hphantom{1}}
\newcommand{\cc}[1]{\multicolumn{1}{c}{#1}}
\renewcommand{\arraystretch}{1.2} % enlarge line spacing
\begin{tabular}{lllll} \hline \hline
& \cc{$\sigma$} & \cc{$A$} & \cc{$C$} & $\chi ^2/N_{\rm DF}$\\ \hline

${\rm 3Q_{Y}}$                 & 
              $0.1524(28)$ & $0.1331(\s 66)$ & $0.9182(213)$ & 3.76\\ 
${\rm 3Q_{Y}}$ \ (Latt. Coul.) & 
              $0.1556(24)$ & $0.1185(\s 53)$ & $0.8876(179)$ & 1.81\\ 
\hline
${\rm Q\bar{Q}}$  (on-axis) & 
              $0.1629(47)$ & $0.2793(116)  $ & $0.6203(161)$ & 0.59\\ 
${\rm Q\bar{Q}}$  (on-axis,  Latt. Coul.) & 
              $0.1603(48)$ & $0.2627(109)  $ & $0.6271(165)$ & 0.51\\ 
${\rm Q\bar{Q}}$  (off-axis, Latt. Coul.) & 
              $0.1611(18)$ & $0.2780(\s 44)  $ & $0.6430(\s 63)$ & 3.57\\ 
\hline\hline
\end{tabular}

$\beta =5.7$ (16 quark configurations)\\

\vspace{1cm}
\begin{tabular}{lllll} \hline \hline
& \cc{$\sigma$} & \cc{$A$} & \cc{$C$} & $\chi ^2/N_{\rm DF}$\\ \hline
${\rm 3Q_{Y}}$                 & 
              $0.1027(\s 6)$ & $0.1230(\s 20)$ & $0.9085(\s 55)$ & 5.03\\ 
${\rm 3Q_{Y}}$ \ (Latt. Coul.) & 
              $0.1031(\s 6)$ & $0.1141(\s 18)$ & $0.8999(\s 54)$ & 4.29\\ 
\hline
${\rm Q\bar{Q}}$  (on-axis) & 
              $0.1079(28)$ & $0.2607(174)  $ & $0.6115(197)$ & 0.92\\ 
${\rm Q\bar{Q}}$  (on-axis,  Latt. Coul.) & 
              $0.1080(28)$ & $0.2377(159)  $ & $0.6074(194)$ & 0.76\\ 
${\rm Q\bar{Q}}$  (off-axis, Latt. Coul.) & 
              $0.1018(11)$ & $0.2795(\s 51)  $ & $0.6596(\s 53)$ & 1.28\\ 
\hline\hline
\end{tabular}

$\beta =5.8$ (139 quark configurations)\\

\vspace{1cm}
\begin{tabular}{lllll} \hline \hline
& \cc{$\sigma$} & \cc{$A$} & \cc{$C$} & $\chi ^2/N_{\rm DF}$\\ \hline
${\rm 3Q_{Y}}$                 & 
              $0.0460(\s 4)$ & $0.1366(\s 11)$ & $0.9599(\s 35)$ & 2.81\\ 
${\rm 3Q_{Y}}$ \ (Latt. Coul.) & 
              $0.0467(\s 4)$ & $0.1256(\s 10)$ & $0.9467(\s 34)$ & 2.22\\ 
\hline
${\rm Q\bar{Q}}$  (on-axis) & 
              $0.0506(\s 7)$ & $0.2768(\s 24)  $ & $0.6374(\s 30)$ & 3.56\\ 
${\rm Q\bar{Q}}$  (on-axis,  Latt. Coul.) & 
              $0.0500(\s 7)$ & $0.2557(\s 22)  $ & $0.6373(\s 30)$ & 1.22\\ 
${\rm Q\bar{Q}}$  (off-axis, Latt. Coul.) & 
              $0.0497(\s 5)$ & $0.2572(\s 15)  $ & $0.6389(\s 20)$ & 1.59\\ 
\hline\hline
\end{tabular}

$\beta =6.0$ (155 quark configurations)\\
\end{center}
\caption
{
The main result on the fit analysis of the lattice QCD data with the Y-ansatz at each $\beta$.
We list the best-fit parameter set $(\sigma, A, C)$ in the function form as 
$V_{\rm 3Q}=-A_{\rm 3Q}\sum_{i<j}\frac{1}{|{\bf r}_i-{\bf r}_j|}
+\sigma_{\rm 3Q}L_{\rm min}+C_{\rm 3Q}$, 
where $L_{\rm min}$ denotes the minimal value of the Y-type flux-tube length.
The label of (Latt. Coul.) means the fit with the lattice Coulomb potential 
instead of the continuum Coulomb potential. 
The similar fit on the Q-$\bar{\rm Q}$ potential is also listed:
``on-axis'' and ``off-axis''  mean the fit analysis only for on-axis data 
and for both on-axis and off-axis data, respectively.
The universality of the string tension  and the OGE result on the Coulomb coefficient are 
found as $\sigma_{\rm 3Q} \simeq \sigma_{\rm Q\bar{\rm Q}}$ and 
$A_{\rm 3Q} \simeq A_{\rm Q\bar{\rm Q}}$, respectively.
The listed values are measured in the lattice unit.
}
\label{fitprm}
\end{table}

\begin{table}[h]
\vspace{0.2cm}
\newcommand{\m}{\hphantom{$-$}}
\newcommand{\cc}[1]{\multicolumn{1}{c}{#1}}
\begin{center}
\begin{tabular}{ c c c c c c c } \hline\hline
$(i, j, k)$ & $V_{\rm 3Q}^{\rm latt}$ & \lower.4ex\hbox{$\bar{C}$} &  
fit range of $T$ & $\chi^2/N_{\rm DF}$
& $V_{\rm 3Q}^{\rm fit}$ & $V^{\rm latt}_{\rm 3Q}-V^{\rm fit}_{\rm 3Q}$ \\
\hline
$(0, 1, 1)$ &0.8457( 38)& 0.9338(173)& 5 --10 & $0.062$ & 0.8524& $-$0.0067 \\ 
$(0, 1, 2)$ &1.0973( 43)& 0.9295(161)& 4 -- 8 & $0.163$ & 1.1025& $-$0.0052 \\ 
$(0, 1, 3)$ &1.2929( 41)& 0.8987(110)& 3 -- 7 & $0.255$ & 1.2929&  \m0.0000 \\
$(0, 2, 2)$ &1.3158( 44)& 0.9151(120)& 3 -- 6 & $0.053$ & 1.3270& $-$0.0112 \\ 
$(0, 2, 3)$ &1.5040( 63)& 0.9041(170)& 3 -- 6 & $0.123$ & 1.5076& $-$0.0036 \\ 
$(0, 3, 3)$ &1.6756( 43)& 0.8718( 73)& 2 -- 5 & $0.572$ & 1.6815& $-$0.0059 \\ 
$(1, 1, 1)$ &1.0238( 40)& 0.9345(149)& 4 -- 8 & $0.369$ & 1.0092&  \m0.0146 \\ 
$(1, 1, 2)$ &1.2185( 62)& 0.9067(228)& 4 -- 8 & $0.352$ & 1.2151&  \m0.0034 \\ 
$(1, 1, 3)$ &1.4161( 49)& 0.9297(135)& 3 -- 7 & $0.842$ & 1.3964&  \m0.0197 \\ 
$(1, 2, 2)$ &1.3866( 48)& 0.9012(127)& 3 -- 7 & $0.215$ & 1.3895& $-$0.0029 \\ 
$(1, 2, 3)$ &1.5594( 63)& 0.8880(165)& 3 -- 6 & $0.068$ & 1.5588&  \m0.0006 \\ 
$(1, 3, 3)$ &1.7145( 43)& 0.8553( 76)& 2 -- 6 & $0.412$ & 1.7202& $-$0.0057 \\ 
$(2, 2, 2)$ &1.5234( 37)& 0.8925( 65)& 2 -- 5 & $0.689$ & 1.5238& $-$0.0004 \\ 
$(2, 2, 3)$ &1.6750(118)& 0.8627(298)& 3 -- 6 & $0.115$ & 1.6763& $-$0.0013 \\ 
$(2, 3, 3)$ &1.8239( 56)& 0.8443( 90)& 2 -- 5 & $0.132$ & 1.8175&  \m0.0064 \\
$(3, 3, 3)$ &1.9607( 93)& 0.8197(154)& 2 -- 5 & $0.000$ & 1.9442&  \m0.0165 \\ 
\hline\hline
\end{tabular}\\[2pt]
\end{center}
\caption{
Lattice QCD results for the 3Q potential $V_{\rm 3Q}^{\rm latt}$ in the lattice unit 
for 16 patterns of the 3Q system at $\beta=5.7$. 
$(i,j,k)$ denotes the 3Q system where the three quarks are 
put on $(i,0,0)$, $(0,j,0)$ and $(0,0,k)$ in ${\bf R}^3$ in the lattice unit.
For each 3Q configuration, $V_{\rm 3Q}^{\rm latt}$ is measured 
from the single-exponential fit as 
$\langle W_{\rm 3Q}\rangle=\bar{C}e^{-V_{\rm 3Q}T}$ 
in the range of $T$ listed at the fourth column. 
The statistical errors listed are estimated with the jackknife method,  
and $\chi^2/N_{\rm DF}$ is listed at the fifth column. 
The best-fit function $V_{\rm 3Q}^{\rm fit}$ in the Y-ansatz is added. 
}
\label{573qpot}
\end{table}

\twocolumn

\begin{table}[h]
\begin{center}
\newcommand{\m}{\hphantom{$-$}}
\newcommand{\cc}[1]{\multicolumn{1}{c}{#1}}
\renewcommand{\tabcolsep}{0.3pc} % enlarge column spacing
\renewcommand{\arraystretch}{1} % enlarge line spacing
\caption{
A part of lattice QCD results for the 3Q potential $V_{\rm 3Q}^{\rm latt}$ at $\beta=5.8$. 
$(i,j,k)$ denotes the 3Q system where the three quarks are 
put on $(i,0,0)$, $(0,j,0)$ and $(0,0,k)$ in ${\bf R}^3$ in the lattice unit.
For each 3Q configuration, $V_{\rm 3Q}^{\rm latt}$ is measured 
from the single-exponential fit as 
$\langle W_{\rm 3Q}\rangle=\bar{C}e^{-V_{\rm 3Q}T}$.
The prefactor $\bar C$ physically means the magnitude of the ground-state component. 
The difference from the best-fit function $V_{\rm 3Q}^{\rm fit}$ in the Y-ansatz is added. 
The listed values are measured in the lattice unit.
}
\label{583qpot1}
\small
\vspace{0.5cm}
\begin{tabular}{c c c c} \hline\hline
$(i, j, k)$ & $V_{\rm 3Q}^{\rm latt}$
& \lower.4ex\hbox{$\bar{C}$} & $V^{\rm latt}_{\rm 3Q}-V^{\rm fit}_{\rm 3Q}$ \\
\hline
 (0, 1, 1 ) &  0.7697( 12) &  0.9554( 58) & $-$0.0041      \\ 
 (0, 1, 2 ) &  0.9639( 28) &  0.9269(128) & $-$0.0039      \\ 
 (0, 1, 3 ) &  1.1112( 60) &  0.9274(274) & \m 0.0053      \\ 
 (0, 1, 4 ) &  1.2337(119) &  0.9106(538) & \m 0.0064      \\ 
 (0, 1, 5 ) &  1.3219(235) &  0.7957(927) & $-$0.0195      \\ 
 (0, 1, 6 ) &  1.4518(132) &  0.8583(442) & \m 0.0000      \\ 
 (0, 1, 7 ) &  1.5719( 72) &  0.8798(182) & \m 0.0119      \\ 
 (0, 1, 8 ) &  1.6621(105) &  0.8189(261) & $-$0.0048      \\ 
 (0, 2, 2 ) &  1.1370( 16) &  0.9342( 46) & $-$0.0018      \\ 
 (0, 2, 3 ) &  1.2659( 21) &  0.9145( 52) & $-$0.0027      \\ 
 (0, 2, 4 ) &  1.3585(239) &  0.7975(946) & $-$0.0279      \\ 
 (0, 2, 5 ) &  1.4834(128) &  0.8448(431) & $-$0.0152      \\ 
 (0, 2, 6 ) &  1.6082( 28) &  0.8810( 50) & \m 0.0004      \\ 
 (0, 2, 8 ) &  1.8251(132) &  0.8548(332) & \m 0.0036      \\ 
 (0, 3, 3 ) &  1.3925( 91) &  0.9168(330) & $-$0.0003      \\ 
 (0, 3, 4 ) &  1.5005( 41) &  0.8862(109) & $-$0.0066      \\ 
 (0, 3, 5 ) &  1.6130( 25) &  0.8810( 41) & $-$0.0042      \\ 
 (0, 3, 6 ) &  1.7171( 32) &  0.8581( 54) & $-$0.0080      \\ 
 (0, 4, 4 ) &  1.6077( 72) &  0.8655(185) & $-$0.0112      \\ 
 (0, 4, 5 ) &  1.7163( 32) &  0.8581( 52) & $-$0.0109      \\ 
 (0, 4, 6 ) &  1.8262( 40) &  0.8482( 67) & $-$0.0075      \\ 
 (0, 4, 7 ) &  1.9321( 51) &  0.8317( 82) & $-$0.0070      \\ 
 (0, 4, 8 ) &  2.0413( 65) &  0.8202(101) & $-$0.0025      \\ 
 (0, 5, 5 ) &  1.8193( 44) &  0.8372( 72) & $-$0.0147      \\ 
 (0, 5, 6 ) &  1.9282( 47) &  0.8265( 76) & $-$0.0111      \\ 
 (0, 5, 8 ) &  2.1460( 73) &  0.8047(113) & $-$0.0017      \\ 
 (0, 6, 6 ) &  2.0322( 62) &  0.8083(102) & $-$0.0113      \\ 
 (0, 6, 7 ) &  2.1401( 71) &  0.7964(112) & $-$0.0070      \\ 
 (0, 6, 8 ) &  2.2384( 85) &  0.7698(123) & $-$0.0118      \\ 
 (0, 7, 7 ) &  2.2461(101) &  0.7813(153) & $-$0.0037      \\ 
 (0, 7, 8 ) &  2.3390(114) &  0.7462(166) & $-$0.0134      \\ 
 (0, 8, 8 ) &  2.4191(177) &  0.6949(244) & $-$0.0351      \\ 
 (1, 1, 1 ) &  0.9140( 32) &  0.9424(147) & \m 0.0149      \\ 
 (1, 1, 2 ) &  1.0647( 42) &  0.9290(194) & \m 0.0096      \\ 
 (1, 1, 3 ) &  1.1914( 86) &  0.8917(384) & \m 0.0053      \\ 
 (1, 1, 4 ) &  1.2879(172) &  0.7887(674) & $-$0.0169      \\ 
 (1, 1, 5 ) &  1.4201( 39) &  0.8662(104) & \m 0.0024      \\ 
 (1, 1, 6 ) &  1.5335( 54) &  0.8600(136) & \m 0.0061      \\ 
 (1, 1, 7 ) &  1.6497( 36) &  0.8686( 59) & \m 0.0146      \\ 
 (1, 1, 8 ) &  1.7557( 46) &  0.8509( 76) & \m 0.0140      \\ 
 (1, 2, 2 ) &  1.1865( 33) &  0.9186(120) & \m 0.0020      \\ 
 (1, 2, 3 ) &  1.3126(124) &  0.9411(576) & \m 0.0072      \\ 
 (1, 2, 4 ) &  1.4155( 28) &  0.8845( 72) & $-$0.0043      \\ 
 (1, 2, 5 ) &  1.5248( 41) &  0.8678(106) & $-$0.0056      \\ 
\hline\hline
\end{tabular}\\[2pt]
\end{center}
\end{table}

\begin{table}[h]
\begin{center}
\newcommand{\m}{\hphantom{$-$}}
\newcommand{\cc}[1]{\multicolumn{1}{c}{#1}}
\renewcommand{\tabcolsep}{0.3pc} % enlarge column spacing
\renewcommand{\arraystretch}{1} % enlarge line spacing
\caption{
A part of lattice QCD results for the 3Q potential $V_{\rm 3Q}^{\rm latt}$ at $\beta=5.8$. 
The notations are the same in Table~\ref{583qpot1}.
}
\label{583qpot2}
\small
\vspace{1cm}
\begin{tabular}{c c c c} \hline\hline
$(i, j, k)$ & $V_{\rm 3Q}^{\rm latt}$
& \lower.4ex\hbox{$\bar{C}$} & $V^{\rm latt}_{\rm 3Q}-V^{\rm fit}_{\rm 3Q}$ \\
\hline
 (1, 2, 6 ) &  1.6356( 57) &  0.8587(145) & $-$0.0031      \\ 
 (1, 2, 7 ) &  1.7490( 36) &  0.8591( 59) & \m 0.0035      \\ 
 (1, 2, 8 ) &  1.8545( 44) &  0.8407( 70) & \m 0.0032      \\ 
 (1, 3, 3 ) &  1.4175( 33) &  0.9020( 86) & $-$0.0024      \\ 
 (1, 3, 4 ) &  1.5301(109) &  0.9083(391) & $-$0.0005      \\ 
 (1, 3, 5 ) &  1.6272( 50) &  0.8538(125) & $-$0.0116      \\ 
 (1, 3, 6 ) &  1.7301( 72) &  0.8285(178) & $-$0.0154      \\ 
 (1, 3, 7 ) &  1.8300(105) &  0.7978(249) & $-$0.0213      \\ 
 (1, 3, 8 ) &  1.9505( 50) &  0.8196( 78) & $-$0.0057      \\ 
 (1, 4, 4 ) &  1.6284( 60) &  0.8652(154) & $-$0.0099      \\ 
 (1, 4, 5 ) &  1.7195( 73) &  0.8133(173) & $-$0.0251      \\ 
 (1, 4, 6 ) &  1.8213( 90) &  0.7883(212) & $-$0.0286      \\ 
 (1, 4, 7 ) &  1.9469( 47) &  0.8216( 72) & $-$0.0075      \\ 
 (1, 4, 8 ) &  2.0543( 59) &  0.8078( 91) & $-$0.0043      \\ 
 (1, 5, 5 ) &  1.8114(119) &  0.7692(271) & $-$0.0378      \\ 
 (1, 5, 7 ) &  2.0496( 55) &  0.8048( 85) & $-$0.0071      \\ 
 (1, 5, 8 ) &  2.1538( 67) &  0.7863(100) & $-$0.0063      \\ 
 (1, 6, 7 ) &  2.1642( 71) &  0.8056(112) & \m 0.0055      \\ 
 (1, 6, 8 ) &  2.2632( 74) &  0.7794(113) & \m 0.0020      \\ 
 (1, 7, 7 ) &  2.2808( 92) &  0.8080(147) & \m 0.0203      \\ 
 (1, 7, 8 ) &  2.3682( 96) &  0.7642(141) & \m 0.0059      \\ 
 (1, 8, 8 ) &  2.4512(156) &  0.7146(217) & $-$0.0123      \\ 
 (2, 2, 2 ) &  1.2771( 73) &  0.9000(258) & $-$0.0041      \\ 
 (2, 2, 3 ) &  1.3783( 80) &  0.8755(277) & $-$0.0107      \\ 
 (2, 2, 4 ) &  1.4899( 42) &  0.8768(107) & $-$0.0074      \\ 
 (2, 2, 5 ) &  1.5933( 54) &  0.8513(133) & $-$0.0113      \\ 
 (2, 2, 8 ) &  1.9188( 49) &  0.8210( 79) & $-$0.0024      \\ 
 (2, 3, 3 ) &  1.4739(120) &  0.8636(410) & $-$0.0142      \\ 
 (2, 3, 4 ) &  1.5831( 22) &  0.8718( 36) & $-$0.0081      \\ 
 (2, 3, 5 ) &  1.6820( 57) &  0.8380(140) & $-$0.0133      \\ 
 (2, 3, 6 ) &  1.7915( 34) &  0.8361( 58) & $-$0.0080      \\ 
 (2, 3, 7 ) &  1.8982( 43) &  0.8220( 68) & $-$0.0053      \\ 
 (2, 3, 8 ) &  2.0002( 53) &  0.7989( 85) & $-$0.0071      \\ 
 (2, 4, 4 ) &  1.6805( 77) &  0.8506(189) & $-$0.0102      \\ 
 (2, 4, 5 ) &  1.7675( 71) &  0.7952(166) & $-$0.0246      \\ 
 (2, 4, 7 ) &  1.9863( 48) &  0.7933( 73) & $-$0.0107      \\ 
 (2, 4, 8 ) &  2.0982( 60) &  0.7875( 94) & $-$0.0015      \\ 
 (2, 5, 8 ) &  2.2004( 72) &  0.7756(106) & \m 0.0051      \\ 
 (2, 6, 6 ) &  2.0932( 65) &  0.7907( 94) & \m 0.0016      \\ 
 (2, 6, 7 ) &  2.1928( 71) &  0.7665(105) & \m 0.0010      \\ 
 (2, 6, 8 ) &  2.2889( 84) &  0.7378(120) & $-$0.0038      \\ 
 (2, 7, 8 ) &  2.4014(106) &  0.7347(147) & \m 0.0103      \\ 
 (3, 3, 3 ) &  1.5566( 72) &  0.8434(180) & $-$0.0197      \\ 
 (3, 3, 4 ) &  1.6474( 66) &  0.8125(160) & $-$0.0253      \\ 
 (3, 3, 5 ) &  1.7641( 37) &  0.8353( 60) & $-$0.0084      \\ 
 (3, 3, 6 ) &  1.8685( 44) &  0.8196( 72) & $-$0.0053      \\ 
 (3, 3, 7 ) &  1.9696( 52) &  0.7965( 82) & $-$0.0062      \\ 
 (3, 3, 8 ) &  2.0753( 65) &  0.7812( 98) & $-$0.0029      \\ 
 (3, 4, 5 ) &  1.8357(106) &  0.7690(240) & $-$0.0249      \\ 
\hline\hline
\end{tabular}\\[2pt]
\end{center}
\end{table}

\begin{table}[h]
\begin{center}
\newcommand{\m}{\hphantom{$-$}}
\newcommand{\cc}[1]{\multicolumn{1}{c}{#1}}
\renewcommand{\tabcolsep}{0.3pc} % enlarge column spacing
\renewcommand{\arraystretch}{1} % enlarge line spacing
\caption{
A part of lattice QCD results for the 3Q potential $V_{\rm 3Q}^{\rm latt}$ at $\beta=5.8$. 
The notations are the same in Table~\ref{583qpot1}.
}
\label{583qpot3}
\small
\vspace{0.5cm}
\begin{tabular}{c c c c} \hline\hline
$(i, j, k)$ & $V_{\rm 3Q}^{\rm latt}$
& \lower.4ex\hbox{$\bar{C}$} & $V^{\rm latt}_{\rm 3Q}-V^{\rm fit}_{\rm 3Q}$ \\
\hline
 (3, 4, 6 ) &  1.9440(157) &  0.7686(358) & $-$0.0153      \\ 
 (3, 4, 7 ) &  2.0569( 60) &  0.7756( 96) & $-$0.0025      \\ 
 (3, 4, 8 ) &  2.1666( 68) &  0.7679(100) & \m 0.0063      \\ 
 (3, 5, 5 ) &  1.9321(178) &  0.7527(394) & $-$0.0220      \\ 
 (3, 5, 6 ) &  2.0504( 56) &  0.7756( 83) & $-$0.0004      \\ 
 (3, 5, 8 ) &  2.2627( 77) &  0.7491(110) & \m 0.0138      \\ 
 (3, 6, 6 ) &  2.1580( 69) &  0.7683(101) & \m 0.0123      \\ 
 (3, 7, 7 ) &  2.3578(110) &  0.7261(155) & \m 0.0190      \\ 
 (3, 7, 8 ) &  2.4496(125) &  0.6934(169) & \m 0.0133      \\ 
 (3, 8, 8 ) &  2.5416(178) &  0.6613(229) & \m 0.0088      \\ 
 (4, 4, 4 ) &  1.8377( 49) &  0.8044( 74) & $-$0.0119      \\ 
 (4, 4, 5 ) &  1.9371( 55) &  0.7900( 82) & $-$0.0044      \\ 
 (4, 4, 6 ) &  2.0367( 61) &  0.7703( 91) & $-$0.0004      \\ 
 (4, 5, 5 ) &  2.0278( 68) &  0.7638( 96) & $-$0.0022      \\ 
 (4, 5, 6 ) &  2.1301( 69) &  0.7503( 99) & \m 0.0073      \\ 
 (4, 6, 6 ) &  2.2310( 87) &  0.7347(123) & \m 0.0174      \\ 
 (4, 6, 8 ) &  2.4356(117) &  0.7003(162) & \m 0.0322      \\ 
 (4, 7, 7 ) &  2.4304(130) &  0.6947(177) & \m 0.0306      \\ 
 (4, 7, 8 ) &  2.5347(150) &  0.6806(201) & \m 0.0402      \\ 
 (4, 8, 8 ) &  2.6412(245) &  0.6696(328) & \m 0.0531      \\ 
 (5, 5, 5 ) &  2.1192( 87) &  0.7417(123) & \m 0.0050      \\ 
 (5, 5, 6 ) &  2.2247( 90) &  0.7334(125) & \m 0.0208      \\ 
 (5, 5, 7 ) &  2.3298(104) &  0.7208(147) & \m 0.0328      \\ 
 (5, 5, 8 ) &  2.4280(131) &  0.6980(182) & \m 0.0357      \\ 
 (5, 6, 6 ) &  2.3159(114) &  0.7065(155) & \m 0.0252      \\ 
 (5, 6, 7 ) &  2.4124(125) &  0.6818(170) & \m 0.0307      \\ 
 (5, 6, 8 ) &  2.5255(151) &  0.6802(200) & \m 0.0503      \\ 
 (5, 8, 8 ) &  2.7193(345) &  0.6369(438) & \m 0.0660      \\ 
 (6, 6, 6 ) &  2.4166(223) &  0.6943(306) & \m 0.0421      \\ 
 (6, 6, 7 ) &  2.5096(190) &  0.6648(252) & \m 0.0468      \\ 
 (6, 6, 8 ) &  2.6408(230) &  0.6868(315) & \m 0.0866      \\ 
 (6, 7, 7 ) &  2.5846(264) &  0.6148(323) & \m 0.0358      \\ 
 (6, 7, 8 ) &  2.7365(260) &  0.6615(343) & \m 0.0981      \\ 
 (6, 8, 8 ) &  2.8742(456) &  0.6939(633) & \m 0.1479      \\ 
\hline\hline
\end{tabular}\\[2pt]
\end{center}
\end{table}

\begin{table}[h]
\begin{center}
\newcommand{\m}{\hphantom{$-$}}
\newcommand{\cc}[1]{\multicolumn{1}{c}{#1}}
\renewcommand{\tabcolsep}{0.3pc} % enlarge column spacing
\renewcommand{\arraystretch}{1} % enlarge line spacing
\caption{
A part of lattice QCD results for the 3Q potential $V_{\rm 3Q}^{\rm latt}$ at $\beta=5.8$. 
$(l,m,n)$ denotes the 3Q system where the three quarks are 
put on $(l,0,0)$, $(-m,0,0)$ and $(0,n,0)$ in ${\bf R}^3$ in the lattice unit.
The other notations are the same in Table 4.
}
\label{583qpot4}
\small
\vspace{0.5cm}
\begin{tabular}{c c c c} \hline\hline
$(l, m, n)$ & $V_{\rm 3Q}^{\rm latt}$
& \lower.4ex\hbox{$\bar{C}$} & $V^{\rm latt}_{\rm 3Q}-V^{\rm fit}_{\rm 3Q}$ \\
\hline
 (1, 2, 1 ) &  1.1185( 22) &  0.9321( 80) & \m 0.0199      \\ 
 (1, 5, 2 ) &  1.5721( 55) &  0.8586(140) & $-$0.0052      \\ 
 (1, 6, 2 ) &  1.6804( 76) &  0.8446(190) & $-$0.0030      \\ 
 (1, 7, 2 ) &  1.7949( 40) &  0.8514( 64) & \m 0.0063      \\ 
 (2, 3, 2 ) &  1.4653( 19) &  0.9028( 33) & \m 0.0068      \\ 
 (2, 4, 2 ) &  1.5757( 21) &  0.8896( 32) & \m 0.0125      \\ 
 (2, 5, 2 ) &  1.6820( 30) &  0.8708( 51) & \m 0.0152      \\ 
 (3, 3, 2 ) &  1.5732( 27) &  0.8879( 45) & \m 0.0144      \\ 
 (3, 4, 2 ) &  1.6793( 33) &  0.8689( 52) & \m 0.0203      \\ 
 (3, 4, 3 ) &  1.7660( 36) &  0.8592( 54) & $-$0.0035      \\ 
 (3, 5, 3 ) &  1.8677( 43) &  0.8377( 67) & $-$0.0002      \\ 
 (4, 4, 3 ) &  1.8692( 48) &  0.8449( 77) & \m 0.0056      \\ 
\hline\hline
\end{tabular}\\[2pt]
\end{center}
\end{table}

\begin{table}[h]
\begin{center}
\newcommand{\m}{\hphantom{$-$}}
\newcommand{\cc}[1]{\multicolumn{1}{c}{#1}}
\renewcommand{\tabcolsep}{0.3pc} % enlarge column spacing
\renewcommand{\arraystretch}{1} % enlarge line spacing
\caption{
A part of lattice QCD results for the 3Q potential $V_{\rm 3Q}^{\rm latt}$ at $\beta=6.0$. 
$(i,j,k)$ denotes the 3Q system where the three quarks are 
put on $(i,0,0)$, $(0,j,0)$ and $(0,0,k)$ in ${\bf R}^3$ in the lattice unit.
For each 3Q configuration, $V_{\rm 3Q}^{\rm latt}$ is measured 
from the single-exponential fit as 
$\langle W_{\rm 3Q}\rangle=\bar{C}e^{-V_{\rm 3Q}T}$.
The prefactor $\bar C$ physically means the magnitude of the ground-state component. 
The difference from the best-fit function $V_{\rm 3Q}^{\rm fit}$ in the Y-ansatz is added. 
The listed values are measured in the lattice unit.
}
\label{603qpot1}
\small
\vspace{0.5cm}
\begin{tabular}{c c c c} \hline\hline
$(i, j, k)$ & $V_{\rm 3Q}^{\rm latt}$
& \lower.4ex\hbox{$\bar{C}$} & $V^{\rm latt}_{\rm 3Q}-V^{\rm fit}_{\rm 3Q}$ \\
\hline
 (0, 1, 1 ) &  0.6778(\m 6) &  0.9784( 24) & $-$0.0012     \\
 (0, 1, 2 ) &  0.8234( 11) &  0.9712( 45) & $-$0.0042      \\
 (0, 1, 3 ) &  0.9183( 17) &  0.9769( 65) & \m 0.0045      \\
 (0, 1, 4 ) &  0.9859( 24) &  0.9589( 92) & \m 0.0050      \\
 (0, 1, 5 ) &  1.0463( 30) &  0.9495(112) & \m 0.0064      \\
 (0, 1, 6 ) &  1.1069( 40) &  0.9595(152) & \m 0.0122      \\
 (0, 1, 7 ) &  1.1572( 50) &  0.9374(192) & \m 0.0102      \\
 (0, 2, 2 ) &  0.9430( 21) &  0.9586( 78) & $-$0.0095      \\
 (0, 2, 3 ) &  1.0259( 24) &  0.9607( 91) & $-$0.0045      \\
 (0, 2, 4 ) &  1.0946( 32) &  0.9657(120) & \m 0.0003      \\
 (0, 2, 5 ) &  1.1454( 41) &  0.9282(149) & $-$0.0064      \\
 (0, 2, 6 ) &  1.2075( 28) &  0.9464( 76) & \m 0.0018      \\
 (0, 2, 7 ) &  1.2563( 33) &  0.9262( 90) & $-$0.0012      \\
 (0, 3, 3 ) &  1.0999( 23) &  0.9566( 62) & $-$0.0031      \\
 (0, 3, 4 ) &  1.1595( 25) &  0.9454( 67) & $-$0.0044      \\
 (0, 3, 5 ) &  1.2170( 25) &  0.9426( 65) & $-$0.0026      \\
 (0, 3, 6 ) &  1.2699( 32) &  0.9327( 90) & $-$0.0027      \\
 (0, 3, 7 ) &  1.3216( 40) &  0.9232(110) & $-$0.0021      \\
 (0, 3, 8 ) &  1.3765( 37) &  0.9241( 92) & \m 0.0029      \\
 (0, 4, 4 ) &  1.2177( 32) &  0.9394( 87) & $-$0.0050      \\
 (0, 4, 5 ) &  1.2723( 34) &  0.9336( 96) & $-$0.0047      \\
 (0, 4, 6 ) &  1.3302( 40) &  0.9418(110) & \m 0.0013      \\
 (0, 4, 7 ) &  1.3744( 42) &  0.9128(108) & $-$0.0048      \\
 (0, 4, 8 ) &  1.4233( 51) &  0.8982(138) & $-$0.0054      \\
 (0, 5, 5 ) &  1.3251( 40) &  0.9265(112) & $-$0.0050      \\
 (0, 5, 6 ) &  1.3762( 39) &  0.9187(108) & $-$0.0049      \\
 (0, 5, 7 ) &  1.4273( 49) &  0.9110(131) & $-$0.0035      \\
 (0, 5, 8 ) &  1.4799( 51) &  0.9079(140) & \m 0.0002      \\
 (0, 6, 6 ) &  1.4248( 52) &  0.9047(136) & $-$0.0066      \\
 (0, 6, 7 ) &  1.4785( 51) &  0.9062(130) & $-$0.0020      \\
 (0, 6, 8 ) &  1.5300( 56) &  0.9020(146) & \m 0.0011      \\
 (0, 7, 7 ) &  1.5314( 35) &  0.9058( 59) & \m 0.0023      \\
 (0, 7, 8 ) &  1.5811( 41) &  0.8971( 66) & \m 0.0039      \\
 (0, 8, 8 ) &  1.6325( 47) &  0.8924( 76) & \m 0.0078      \\
 (1, 1, 1 ) &  0.7900( 21) &  0.9588( 98) & \m 0.0073      \\
 (1, 1, 2 ) &  0.8992( 25) &  0.9707(118) & \m 0.0044      \\
 (1, 1, 3 ) &  0.9800( 38) &  0.9578(182) & \m 0.0052      \\
 (1, 1, 4 ) &  1.0515( 25) &  0.9677( 99) & \m 0.0115      \\
 (1, 1, 5 ) &  1.1105( 36) &  0.9578(135) & \m 0.0123      \\
 (1, 1, 6 ) &  1.1645( 47) &  0.9449(175) & \m 0.0120      \\
 (1, 1, 7 ) &  1.2140( 59) &  0.9227(209) & \m 0.0095      \\
 (1, 1, 8 ) &  1.2662( 72) &  0.9146(260) & \m 0.0111      \\
 (1, 2, 2 ) &  0.9796( 18) &  0.9600( 68) & $-$0.0019      \\
 (1, 2, 3 ) &  1.0555( 23) &  0.9743( 89) & \m 0.0029      \\
\hline\hline
\end{tabular}\\[2pt]
\end{center}
\end{table}

\begin{table}[h]
\begin{center}
\newcommand{\m}{\hphantom{$-$}}
\newcommand{\cc}[1]{\multicolumn{1}{c}{#1}}
\renewcommand{\tabcolsep}{0.3pc} % enlarge column spacing
\renewcommand{\arraystretch}{1} % enlarge line spacing
\caption{
A part of lattice QCD results for the 3Q potential $V_{\rm 3Q}^{\rm latt}$ at $\beta=6.0$. 
The notations are the same in Table~\ref{603qpot1}.
}
\label{603qpot2}
\small
\vspace{1cm}
\begin{tabular}{c c c c} \hline\hline
$(i, j, k)$ & $V_{\rm 3Q}^{\rm latt}$
& \lower.4ex\hbox{$\bar{C}$} & $V^{\rm latt}_{\rm 3Q}-V^{\rm fit}_{\rm 3Q}$ \\
\hline
 (1, 2, 4 ) &  1.1185( 31) &  0.9689(119) & \m 0.0043      \\
 (1, 2, 5 ) &  1.1720( 39) &  0.9458(142) & \m 0.0013      \\
 (1, 2, 6 ) &  1.2283( 49) &  0.9467(182) & \m 0.0043      \\
 (1, 2, 7 ) &  1.2712( 60) &  0.9015(208) & $-$0.0042      \\
 (1, 2, 8 ) &  1.3314( 66) &  0.9262(243) & \m 0.0058      \\
 (1, 3, 3 ) &  1.1166( 34) &  0.9582(132) & $-$0.0013      \\ 
 (1, 3, 4 ) &  1.1783( 35) &  0.9659(137) & \m 0.0018      \\ 
 (1, 3, 5 ) &  1.2299( 45) &  0.9453(174) & $-$0.0011      \\ 
 (1, 3, 6 ) &  1.2877( 32) &  0.9513( 87) & \m 0.0045      \\ 
 (1, 3, 7 ) &  1.3293( 58) &  0.9095(210) & $-$0.0047      \\ 
 (1, 3, 8 ) &  1.3863( 40) &  0.9214(103) & \m 0.0027      \\ 
 (1, 4, 4 ) &  1.2296( 29) &  0.9442( 75) & $-$0.0030      \\ 
 (1, 4, 5 ) &  1.2863( 29) &  0.9484( 78) & \m 0.0006      \\ 
 (1, 4, 6 ) &  1.3326( 35) &  0.9236( 90) & $-$0.0043      \\ 
 (1, 4, 7 ) &  1.3864( 40) &  0.9249(102) & $-$0.0004      \\ 
 (1, 4, 8 ) &  1.4401( 44) &  0.9249(117) & \m 0.0042      \\ 
 (1, 5, 5 ) &  1.3363( 37) &  0.9370( 99) & $-$0.0013      \\ 
 (1, 5, 6 ) &  1.3879( 37) &  0.9316( 99) & \m 0.0001      \\ 
 (1, 5, 7 ) &  1.4318( 44) &  0.9039(116) & $-$0.0053      \\ 
 (1, 5, 8 ) &  1.4837( 47) &  0.9001(127) & $-$0.0020      \\ 
 (1, 6, 6 ) &  1.4344( 52) &  0.9115(138) & $-$0.0029      \\ 
 (1, 6, 7 ) &  1.4870( 55) &  0.9115(150) & \m 0.0009      \\ 
 (1, 6, 8 ) &  1.5370( 53) &  0.9037(142) & \m 0.0030      \\ 
 (1, 7, 7 ) &  1.5258( 65) &  0.8729(165) & $-$0.0084      \\ 
 (1, 7, 8 ) &  1.5743( 69) &  0.8624(168) & $-$0.0075      \\ 
 (1, 8, 8 ) &  1.6370( 44) &  0.8888( 69) & \m 0.0080      \\ 
 (2, 2, 2 ) &  1.0405( 34) &  0.9669(132) & \m 0.0004      \\ 
 (2, 2, 3 ) &  1.0963( 31) &  0.9462(115) & $-$0.0045      \\ 
 (2, 2, 4 ) &  1.1579( 37) &  0.9570(144) & $-$0.0002      \\ 
 (2, 2, 5 ) &  1.2108( 49) &  0.9396(181) & $-$0.0016      \\ 
 (2, 2, 6 ) &  1.2676( 32) &  0.9446( 90) & \m 0.0031      \\ 
 (2, 2, 7 ) &  1.3035( 69) &  0.8803(246) & $-$0.0117      \\ 
 (2, 2, 8 ) &  1.3613( 42) &  0.9000(112) & $-$0.0035      \\ 
 (2, 3, 3 ) &  1.1461( 39) &  0.9322(147) & $-$0.0087      \\ 
 (2, 3, 4 ) &  1.1994( 40) &  0.9247(149) & $-$0.0089      \\ 
 (2, 3, 5 ) &  1.2525( 52) &  0.9200(188) & $-$0.0079      \\ 
 (2, 3, 6 ) &  1.3114( 32) &  0.9368( 85) & \m 0.0002      \\ 
 (2, 3, 7 ) &  1.3499( 71) &  0.8852(256) & $-$0.0110      \\ 
 (2, 3, 8 ) &  1.4000( 79) &  0.8775(280) & $-$0.0099      \\ 
 (2, 4, 4 ) &  1.2565( 28) &  0.9443( 80) & $-$0.0027      \\ 
 (2, 4, 5 ) &  1.3041( 34) &  0.9274( 89) & $-$0.0053      \\ 
 (2, 4, 6 ) &  1.3549( 36) &  0.9207(100) & $-$0.0041      \\ 
 (2, 4, 7 ) &  1.4006( 37) &  0.8990( 98) & $-$0.0072      \\ 
 (2, 4, 8 ) &  1.4535( 49) &  0.8980(133) & $-$0.0026      \\ 
 (2, 5, 5 ) &  1.3505( 42) &  0.9129(111) & $-$0.0077      \\ 
 (2, 5, 6 ) &  1.4074( 42) &  0.9272(116) & \m 0.0007      \\ 
 (2, 5, 7 ) &  1.4520( 48) &  0.9040(126) & $-$0.0028      \\ 
 (2, 6, 6 ) &  1.4515( 49) &  0.9055(130) & $-$0.0029      \\ 
 (2, 6, 7 ) &  1.4920( 53) &  0.8722(135) & $-$0.0098      \\ 
\hline\hline
\end{tabular}\\[2pt]
\end{center}
\end{table}

\begin{table}[h]
\begin{center}
\newcommand{\m}{\hphantom{$-$}}
\newcommand{\cc}[1]{\multicolumn{1}{c}{#1}}
\renewcommand{\tabcolsep}{0.3pc} % enlarge column spacing
\renewcommand{\arraystretch}{1} % enlarge line spacing
\caption{
A part of lattice QCD results for the 3Q potential $V_{\rm 3Q}^{\rm latt}$ at $\beta=6.0$. 
The notations are the same in Table~\ref{603qpot1}.
}
\label{603qpot3}
\small
\vspace{0.5cm}
\begin{tabular}{c c c c} \hline\hline
$(i, j, k)$ & $V_{\rm 3Q}^{\rm latt}$
& \lower.4ex\hbox{$\bar{C}$} & $V^{\rm latt}_{\rm 3Q}-V^{\rm fit}_{\rm 3Q}$ \\
\hline
 (2, 6, 8 ) &  1.5528( 55) &  0.8950(133) & \m 0.0039      \\ 
 (2, 7, 7 ) &  1.5442( 61) &  0.8727(163) & $-$0.0044      \\ 
 (2, 7, 8 ) &  1.5910( 71) &  0.8585(177) & $-$0.0042      \\ 
 (2, 8, 8 ) &  1.6411( 84) &  0.8518(213) & $-$0.0004      \\ 
 (3, 3, 3 ) &  1.1951( 31) &  0.9424( 86) & $-$0.0060      \\ 
 (3, 3, 4 ) &  1.2421( 28) &  0.9308( 77) & $-$0.0080      \\ 
 (3, 3, 5 ) &  1.2936( 34) &  0.9280( 93) & $-$0.0059      \\ 
 (3, 3, 6 ) &  1.3404( 42) &  0.9075(111) & $-$0.0082      \\ 
 (3, 3, 8 ) &  1.4434( 52) &  0.8992(141) & $-$0.0020      \\ 
 (3, 4, 4 ) &  1.2894( 32) &  0.9292( 79) & $-$0.0066      \\ 
 (3, 4, 5 ) &  1.3350( 34) &  0.9142( 88) & $-$0.0081      \\ 
 (3, 4, 6 ) &  1.3825( 40) &  0.9014(103) & $-$0.0082      \\ 
 (3, 4, 7 ) &  1.4353( 49) &  0.9033(131) & $-$0.0029      \\ 
 (3, 4, 8 ) &  1.4867( 56) &  0.8987(145) & \m 0.0011      \\ 
 (3, 5, 6 ) &  1.4273( 42) &  0.8929(114) & $-$0.0075      \\ 
 (3, 5, 7 ) &  1.4791( 53) &  0.8931(138) & $-$0.0023      \\ 
 (3, 5, 8 ) &  1.5260( 50) &  0.8770(129) & $-$0.0020      \\ 
 (3, 6, 6 ) &  1.4737( 61) &  0.8812(162) & $-$0.0064      \\ 
 (3, 6, 7 ) &  1.5234( 51) &  0.8769(124) & $-$0.0024      \\ 
 (3, 6, 8 ) &  1.5734( 61) &  0.8709(147) & \m 0.0015      \\ 
 (3, 7, 7 ) &  1.5655( 77) &  0.8535(203) & $-$0.0054      \\ 
 (3, 7, 8 ) &  1.6177( 77) &  0.8550(189) & \m 0.0013      \\ 
 (3, 8, 8 ) &  1.6666( 92) &  0.8458(222) & \m 0.0054      \\ 
 (4, 4, 4 ) &  1.3255( 46) &  0.9047(119) & $-$0.0126      \\ 
 (4, 4, 5 ) &  1.3765( 41) &  0.9116(111) & $-$0.0061      \\ 
 (4, 4, 6 ) &  1.4262( 49) &  0.9088(129) & $-$0.0021      \\ 
 (4, 4, 7 ) &  1.4745( 60) &  0.8985(155) & \m 0.0000      \\ 
 (4, 4, 8 ) &  1.5243( 69) &  0.8925(174) & \m 0.0034      \\ 
 (4, 5, 5 ) &  1.4165( 49) &  0.8903(132) & $-$0.0086      \\ 
 (4, 5, 6 ) &  1.4628( 51) &  0.8808(124) & $-$0.0065      \\ 
 (4, 5, 7 ) &  1.5155( 59) &  0.8868(155) & \m 0.0012      \\ 
 (4, 5, 8 ) &  1.5651( 63) &  0.8786(162) & \m 0.0053      \\ 
 (4, 6, 6 ) &  1.5198( 65) &  0.9031(175) & \m 0.0076      \\ 
 (4, 6, 7 ) &  1.5549( 66) &  0.8608(163) & $-$0.0013      \\ 
 (4, 6, 8 ) &  1.6045( 75) &  0.8552(188) & \m 0.0036      \\ 
 (5, 5, 5 ) &  1.4653( 75) &  0.8984(203) & \m 0.0000      \\ 
 (5, 5, 6 ) &  1.5074( 63) &  0.8814(158) & $-$0.0003      \\ 
 (5, 5, 7 ) &  1.5452( 80) &  0.8475(199) & $-$0.0061      \\ 
 (5, 5, 8 ) &  1.6008( 83) &  0.8574(208) & \m 0.0050      \\ 
 (5, 6, 6 ) &  1.5511( 72) &  0.8689(181) & \m 0.0026      \\ 
 (5, 6, 8 ) &  1.6385( 81) &  0.8343(195) & \m 0.0040      \\ 
 (5, 7, 7 ) &  1.6354( 89) &  0.8288(216) & \m 0.0029      \\ 
 (5, 7, 8 ) &  1.6878( 93) &  0.8325(219) & \m 0.0127      \\ 
 (6, 6, 6 ) &  1.5953(108) &  0.8605(272) & \m 0.0076      \\ 
 (6, 6, 7 ) &  1.6328( 91) &  0.8285(222) & \m 0.0040      \\ 
 (6, 6, 8 ) &  1.6956( 46) &  0.8595( 67) & \m 0.0244      \\ 
 (6, 8, 8 ) &  1.7701(138) &  0.7972(325) & \m 0.0193      \\ 
 (7, 7, 8 ) &  1.7666(139) &  0.7915(326) & \m 0.0188      \\ 
 (7, 8, 8 ) &  1.8114(166) &  0.7795(376) & \m 0.0243      \\ 
\hline\hline
\end{tabular}\\[2pt]
\end{center}
\end{table}

\begin{table}[h]
\begin{center}
\newcommand{\m}{\hphantom{$-$}}
\newcommand{\cc}[1]{\multicolumn{1}{c}{#1}}
\renewcommand{\tabcolsep}{0.3pc} % enlarge column spacing
\renewcommand{\arraystretch}{1} % enlarge line spacing
\caption{
A part of lattice QCD results for the 3Q potential $V_{\rm 3Q}^{\rm latt}$ at $\beta=6.0$. 
$(l,m,n)$ denotes the 3Q system where the three quarks are 
put on $(l,0,0)$, $(-m,0,0)$ and $(0,n,0)$ in ${\bf R}^3$ in the lattice unit.
The other notations are the same in Table 8.
}
\label{603qpot4}
\small
\vspace{0.5cm}
\begin{tabular}{c c c c} \hline\hline
$(l, m, n)$ & $V_{\rm 3Q}^{\rm latt}$
& \lower.4ex\hbox{$\bar{C}$} & $V^{\rm latt}_{\rm 3Q}-V^{\rm fit}_{\rm 3Q}$ \\
\hline
 (1, 2, 1 ) &  0.9334( 16) &  0.9726( 64) & \m 0.0098      \\ 
 (1, 5, 2 ) &  1.2001( 26) &  0.9529( 74) & \m 0.0061      \\ 
 (1, 6, 2 ) &  1.2467( 58) &  0.9211(214) & \m 0.0009      \\ 
 (2, 3, 2 ) &  1.1436( 22) &  0.9600( 61) & \m 0.0053      \\ 
 (2, 4, 2 ) &  1.2017( 26) &  0.9555( 72) & \m 0.0094      \\ 
 (2, 5, 2 ) &  1.2543( 31) &  0.9418( 84) & \m 0.0106      \\ 
 (2, 6, 2 ) &  1.3033( 37) &  0.9218(105) & \m 0.0097      \\ 
 (2, 6, 3 ) &  1.3496( 44) &  0.9202(120) & $-$0.0007      \\ 
 (3, 3, 2 ) &  1.2022( 27) &  0.9537( 77) & \m 0.0101      \\ 
 (3, 4, 2 ) &  1.2563( 32) &  0.9405( 90) & \m 0.0130      \\ 
 (3, 4, 3 ) &  1.2971( 31) &  0.9330( 86) & $-$0.0011      \\ 
 (3, 5, 3 ) &  1.3467( 40) &  0.9215(108) & \m 0.0006      \\ 
 (4, 4, 3 ) &  1.3440( 44) &  0.9147(117) & $-$0.0008      \\ 
\hline\hline
\end{tabular}\\[2pt]
\end{center}
\end{table}

\onecolumn

\begin{table}[h]
\begin{center}
\newcommand{\m}{\hphantom{$-$}}
\newcommand{\s}{\hphantom{1}}
\newcommand{\cc}[1]{\multicolumn{1}{c}{#1}}
\renewcommand{\arraystretch}{1.2} % enlarge line spacing
\begin{tabular}{llllll} \hline \hline
& \cc{$\beta$} & \cc{$\sigma_\Delta$} & \cc{$A_\Delta$} & \cc{$C_\Delta$} & $\chi^2/N_{\rm DF}$\\
\hline
${\rm 3Q_{\Delta}}$                 & 5.7 &
              $0.0858(16)$ & $0.1410(\s 64)$ & $0.9334(210)$ & 10.1\\ 
${\rm 3Q_{\Delta}}$ \ (Latt. Coul.) &     &
              $0.0868(14)$ & $0.1296(\s 51)$ & $0.9146(173)$ & 5.12\\ 
\hline
${\rm 3Q_{\Delta}}$                 & 5.8 &
              $0.0581(\s 4)$ & $0.1197(\s 19)$ & $0.8964(\s 55)$ & 13.7\\ 
${\rm 3Q_{\Delta}}$ \ (Latt. Coul.) &     &
              $0.0583(\s 3)$ & $0.1110(\s 18)$ & $0.8872(\s 54)$ & 13.6\\ 
\hline
${\rm 3Q_{\Delta}}$                 & 6.0 &
              $0.0264(\s 2)$ & $0.1334(\s 11)$ & $0.9490(\s 36)$ & 3.74\\ 
${\rm 3Q_{\Delta}}$ \ (Latt. Coul.) &     &
              $0.0268(\s 2)$ & $0.1227(\s 10)$ & $0.9361(\s 34)$ & 2.89\\ 
\hline\hline
\end{tabular}
\end{center}
\caption{
The fit analysis of the lattice QCD data $V^{\rm latt}_{\rm 3Q}$ with the $\Delta$-ansatz at each $\beta$.
We list the best-fit parameter set 
$(\sigma_\Delta, A_\Delta, C_\Delta)$ in the function form as 
$-A_\Delta \sum_{i<j}\frac{1}{|{\bf r}_i-{\bf r}_j|}+
\sigma_\Delta \sum_{i<j}|{\bf r}_i-{\bf r}_j|+C_\Delta$.
The label of (Latt. Coul.) means the fit with the lattice Coulomb potential 
instead of the continuum Coulomb potential. 
The listed values are measured in the lattice unit.}
\label{dfitprm}
\end{table}

\normalsize
\begin{table}[h]
\begin{center}
\newcommand{\m}{\hphantom{$-$}}
\newcommand{\s}{\hphantom{1}}
\newcommand{\cc}[1]{\multicolumn{1}{c}{#1}}
\renewcommand{\arraystretch}{1.2} % enlarge line spacing
\begin{tabular}{ccccccc} \hline \hline
\cc{$\beta$} & \cc{$\sigma_{\rm GY}$} & \cc{$A_{\rm GY}$} & \cc{$C_{\rm GY}$} & 
$R_{\rm core}$ [$a$] & $R_{\rm core}$ [fm] & $\chi ^2/N_{\rm DF}$
\\ \hline
5.8 & $0.1054(\s 6)$ & $0.1354(\s 18)$ & $0.9569(\s 53)$ & $0.57$ 
& $0.08$ &2.63\\ 
6.0 & $0.0480(\s 4)$ & $0.1451(\s 11)$ & $0.9837(\s 33)$ & $0.79$ 
& $0.08$ &1.23\\ 
\hline\hline
\end{tabular}
\end{center}
\caption{
The fit analysis of the lattice QCD data $V^{\rm latt}_{\rm 3Q}$ with the generalized Y-ansatz at each $\beta$.
We list the best-fit parameter set $(\sigma_{\rm GY}, A_{\rm GY}, C_{\rm GY}, R_{\rm core})$ 
in the lattice unit at $\beta$=5.8 and 6.0. The flux-tube core radius $R_{\rm core}$ in 
the physical unit is added.
}
\label{gyfitprm}
\end{table}

\end{document}